\documentclass[fleqn,10pt]{wlscirep}

\usepackage[utf8]{inputenc}
\usepackage[T1]{fontenc}

\usepackage{subcaption}
\usepackage{xr}
\usepackage{verbatim}
\usepackage{adjustbox}
\usepackage{xcolor}
\usepackage{tabu}
\usepackage{tabularx}
\usepackage{multirow}

\usepackage[style=nature,backend=bibtex]{biblatex}
\let\cite=\supercite

\newcommand{\sflabel}[1]{(\textbf{#1})}

\title{The h-index is no longer an effective correlate of scientific reputation}

\author[1,*]{Vladlen Koltun}
\author[2]{David Hafner}

\affil[1]{Intelligent Systems Lab, Intel, Santa Clara, CA 95054, USA}
\affil[2]{Intelligent Systems Lab, Intel, 85579 Neubiberg, Germany}

\affil[*]{vladlen.koltun@intel.com}

\keywords{research metrics, science of science}

\begin{abstract}
    The impact of individual scientists is commonly quantified using citation-based measures. The most common such measure is the h-index. A scientist's h-index affects hiring, promotion, and funding decisions, and thus shapes the progress of science. Here we report a large-scale study of scientometric measures, analyzing millions of articles and hundreds of millions of citations across four scientific fields and two data platforms. We find that the correlation of the h-index with awards that indicate recognition by the scientific community has substantially declined. These trends are associated with changing authorship patterns. We show that these declines can be mitigated by fractional allocation of citations among authors, which has been discussed in the literature but not implemented at scale. We find that a fractional analogue of the h-index outperforms other measures as a correlate and predictor of scientific awards. Our results suggest that the use of the h-index in ranking scientists should be reconsidered, and that fractional allocation measures such as h-frac provide more robust alternatives.

\end{abstract}

\begin{document}

\flushbottom
\maketitle

\thispagestyle{empty}

\begin{refsegment}
    \defbibfilter{notother}{not segment=\therefsegment}

    \section*{Introduction}

The h-index, proposed by Hirsch in 2005~\cite{Hirsch2005}, has become the leading measure for quantifying the impact of a scientist's published work. The h-index is prominently featured in citation databases such as Google Scholar, Scopus, and Web of Science. It informs hiring, promotion, and funding decisions~\cite{Abbott2010,McNutt2014,Hicks2015}. It thereby shapes the evolution of the scientific community and the progress of science.

Numerous variants of the h-index have been explored, and sophisticated alternatives have been proposed~\cite{PanaretosMalesios2009,Sinatra2016}. None of these has displaced the h-index as the dominant measure of a scientist's output. The endurance of the h-index can be attributed to a number of characteristics. First, it summarizes a scientist's output in a single number that can be readily used for comparison and ranking. Second, it does not require a minimal number of publications or career length, and can thus be computed for scientists at all career stages. Third, it does not require tuning thresholds or parameters. Fourth, it is easily interpretable. Lastly, criticism notwithstanding, the h-index is seen as a robust measure of an individual scientist's impact~\cite{Hirsch2007,Radicchi2008,Henzinger2010,Acuna2012}.

Science continues to evolve and publication patterns change over time~\cite{Fortunato2018}.
Here we report an extensive empirical evaluation of individual research metrics.
Since publication patterns differ across scientific fields~\cite{Alonso2009,Waltman2015,Waltman2016},
we collect large datasets in four fields of research: biology, computer science, economics, and physics.
In each field, we consider 1,000 most highly cited researchers and trace their published output and its impact through two bibliographic data platforms:
Scopus~\cite{scopus} and Google Scholar~\cite{scholar}. The resulting datasets comprise 1.3 million articles and 102 million citations identified via Scopus and 2.6 million articles and 221 million citations identified via Google Scholar (Supplementary Fig.~\ref{fig:data_cum}).

We have cross-referenced the scientists in our datasets against lists of recipients of scientific awards that indicate recognition by the scientific community: Nobel Prizes, Breakthrough Prizes, membership in the National Academies, fellowship of the American Physical Society, Turing Award, fellowship of the Econometric Society, and other distinctions (Supplementary Fig.~\ref{fig:awards} and Supplementary Table~\ref{tab:awards}).
Among the 4,000 authors in our dataset, 75.6\% have no such awards, 13.3\% have one award, 5.1\% have two, and 6.0\% have three or more (Supplementary Fig.~\ref{fig:awards_dist}).
Our basic methodology is to correlate rankings induced by scientometric measures with rankings induced by scientific awards. The assumption is that a citation-based measure that more reliably uncovers laureates of elite awards is a more veridical indicator of scientific reputation~\cite{Sinatra2016,Ioannidis2016}.
Since publication, citation, and award patterns differ substantially across fields, we conduct parallel experiments in the four fields of research. To confirm the robustness of the findings, the studies are replicated across the two bibliographic platforms (Scopus and Google Scholar).

A number of prior studies are related to our work.
Sinatra et al.~\cite{Sinatra2016} analyze the careers of 2,887 physicists in the APS dataset and 7,630 scientists in the Web of Science database, considering approximately one million publications in total. Their study includes evaluations that correlate individual scientific impact indicators with scientific awards. However, this is performed on a limited scale, taking into account only Nobel prizes in physics and Dirac and Boltzmann medals as indicators of scientific reputation.
Considering publication and citation data of 84,116 scientists, Ioannidis et al.~\cite{Ioannidis2016} investigate a number of citation indicators based on how well they capture Nobel prize winners from the years 2011--2015.
The recent study of Ayaz and Masood~\cite{Ayaz2020} evaluates indices of researchers' impact by analyzing 236,416 publications in the area of computer science. Their comparison of bibliometric indices is based on 47 award winners in their dataset.

Our study is conducted on a much larger scale. We analyze millions of articles in four different research fields that are cited hundreds of millions of times. We collect more than 10,000 awards and trace 1,848 distinct awards to the 4,000 scientists in our dataset. (See supplementary information.)
Most importantly, our datasets have yearly temporal granularity from 1970 onwards. This enables detailed evaluation of the temporal evolution of the effectiveness and predictive power of research metrics that, to the best of our knowledge, has not been presented before.

Our first major finding is that the effectiveness of scientometric measures is declining. For example, the correlation of the h-index with scientific awards in physics has dropped from 0.34 in 2010 to 0.00 in 2019 (Kendall's $\tau$, Scopus physics dataset). This is associated with changing authorship patterns, including a higher prevalence of hyperauthorship. Our second major finding is that fractional allocation of citations among coauthors can mitigate this decline~\cite{Price1981,Egghe2008,Waltman2016}. In particular, for each measure we study, its fractional counterpart is a better correlate and predictor of scientific awards. Among all measures, a fractional analogue of the h-index, h-frac, consistently outperforms alternatives.

We test the robustness of the findings via controlled experiments across datasets. The main findings hold in all conditions: fractional allocation improves the effectiveness and predictive power of research metrics, and h-frac is consistently the most reliable bibliometric indicator.
Our results suggest that the use of the h-index in ranking scientists should be reconsidered, and that fractional allocation measures such as h-frac provide more robust alternatives.
The data also indicate, contrary to concerns expressed in the literature, that fractional allocation measures are not antithetic to collaboration.
Our findings can lead to more effective distribution of resources and thus accelerate scientific discovery~\cite{Ioannidis2015}.
Our data, methodology, and findings may also have broader applications in the empirical analysis of science~\cite{Fortunato2018}.

\section*{Results}

\subsection*{Declining effectiveness of individual research metrics}

\begin{figure*}[t!]
	\centering
	\begin{subfigure}[t]{0.5\textwidth}
		\centering
        \caption{}
        \label{fig:effectiveness_scopus_physics}
		\includegraphics[width=\textwidth]{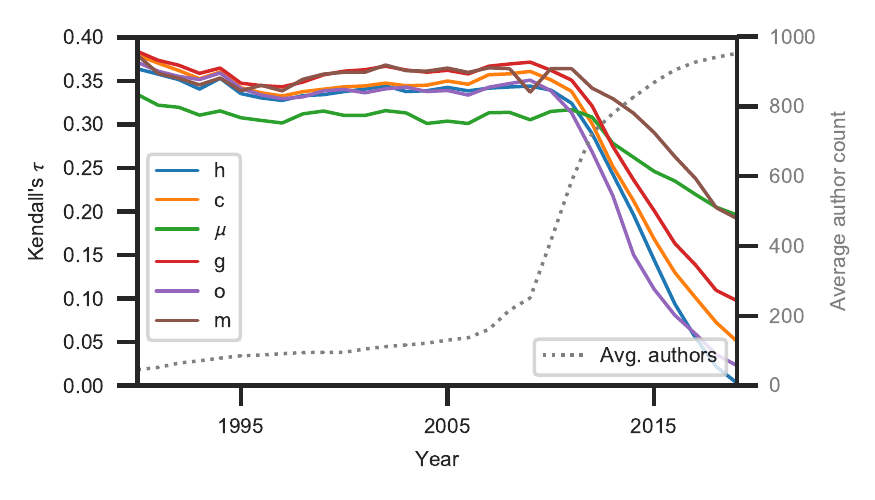}
	\end{subfigure}%
	\begin{subfigure}[t]{0.5\textwidth}
		\centering
        \caption{}
        \label{fig:num_authors_scopus_physics}
		\includegraphics[width=\textwidth]{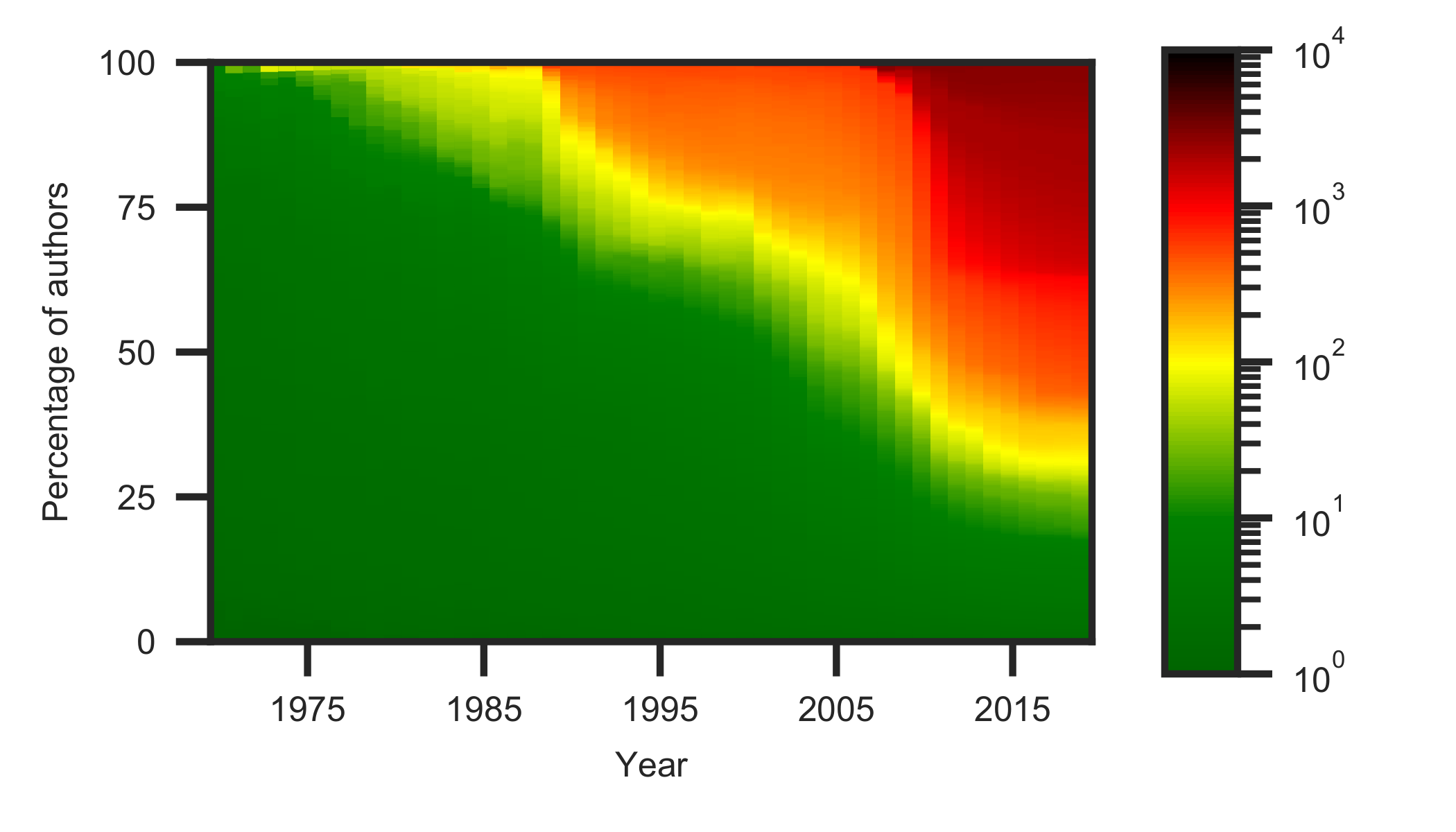}
	\end{subfigure}\\[-5mm]
	\begin{subfigure}[t]{\textwidth}
		\centering
        \caption{}
        \label{fig:evolution_scopus_physics}
		\includegraphics[width=\textwidth]{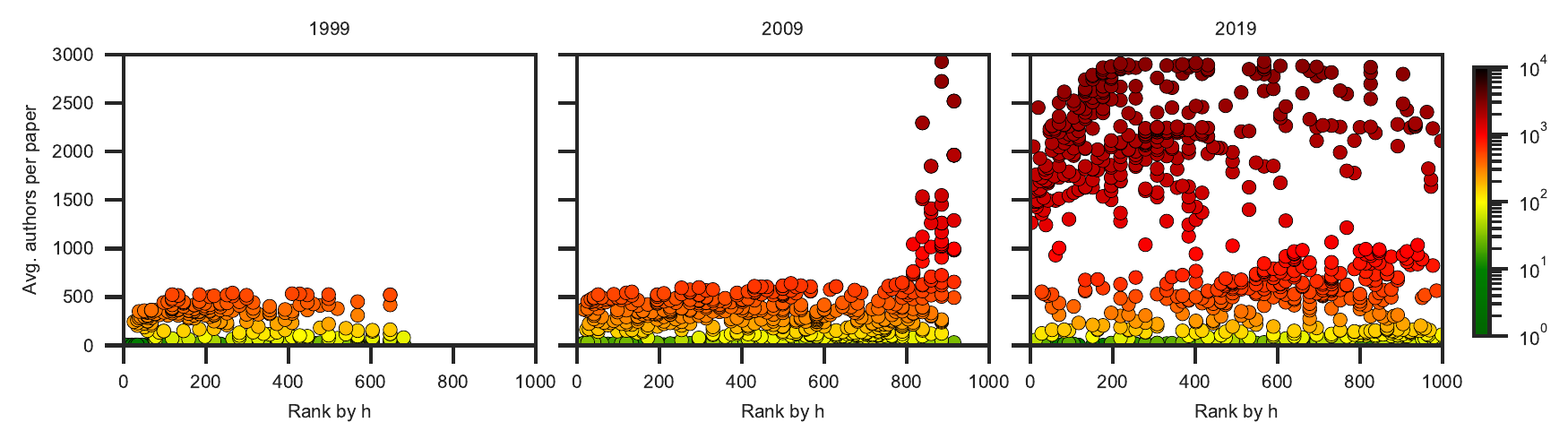}
	\end{subfigure}%
	\caption{The effectiveness of scientometric measures is declining.
        \sflabel{a}~Effectiveness of scientometric measures as correlates of scientific awards in the Scopus physics dataset.
        \sflabel{b}~Color-coded distribution of the average number of coauthors per publication in this dataset.
        \sflabel{c}~Ranking of physicists by the h-index. Each data point is a scientist. Color and the vertical axis represent the average number of coauthors per publication.
	}
\end{figure*}

Fig.~\ref{fig:effectiveness_scopus_physics} shows the effectiveness of scientometric measures over the past 30 years. The effectiveness of a scientometric measure is quantified by the correlation between the ranking induced by this measure and the ranking induced by scientific community awards at a given point in time. Here we report Kendall's $\tau$ on the Scopus physics dataset (see Supplementary Fig.~\ref{fig:criteria_all} for other correlation criteria and datasets). In addition to the h-index~(h), we evaluate the total number of citations to a scientist's work~(c), the mean number of citations per paper~($\mu$, advocated by Lehmann et al.~\cite{Lehmann2006}), Egghe's g-index~\cite{Egghe2006}, the o-index~\cite{DorogovtsevMendes2015}, and the median number of citations received by a scientist's highly-cited papers~(m, highlighted by Bornmann et al.~\cite{Bornmann2008}). (See supplementary information.)

As Fig.~\ref{fig:effectiveness_scopus_physics} demonstrates, the effectiveness of scientometric measures has declined. The decline is particularly pronounced for the h-index.
The effectiveness of the h-index, as measured by Kendall's $\tau$, varied between 0.33 and 0.36 from 1990 to 2010, but dropped to 0.00 by 2019 on the Scopus physics dataset. This is concomitant with a dramatic shift in authorship patterns, illustrated by the average number of coauthors per paper for highly-cited physicists. While the mean number of coauthors per publication, averaged across highly cited physicists, was 78 in 1994 and 121 in 2004,
it rose to 952 in 2019, with 10\% of the scientists having more than 2,441 coauthors per publication on average. (See \url{https://h-frac.org/dataset-s2}.)

This is further illustrated in Fig.~\ref{fig:num_authors_scopus_physics}, which shows the distribution of the average number of coauthors per paper for highly-cited physicists in each year from 1970 onwards. While small authorship teams were nearly universal in the beginning of this period (84\% of the scientists had $<$10 coauthors per publication on average in 1980), the set of highly-cited physicists has come to be dominated by ``hyper-collaborators'': 68\% of the scientists had $>$100 coauthors per publication on average in 2019. Large-scale collaboration has been a feature of science for centuries, but joint authorship has been institutionalized on a new scale in the past decade~\cite{Ioannidis2016}. Scientific consortia comprise thousands of authors who jointly author hundreds of publications~\cite{King2012}. All members of the consortium are listed as authors on all papers~\cite{ATLAS2010}.
This has been referred to as hyperauthorship~\cite{Castelvecchi2015,Milojevic2014}. Our results indicate that this behavior is reducing the effectiveness of established scientometric indicators. This is further illustrated in Fig.~\ref{fig:evolution_scopus_physics}, which shows the ranking of physicists by h-index in 1999, 2009, and 2019. The hyper-collaborators have permeated the ranking.

\subsection*{Fractional allocation}

Are there scientific impact metrics that share the advantages of the h-index and are robust to contemporary publication patterns? Hirsch proposed a bibliometric indicator that takes authorship into account~\cite{Hirsch2010}, but his mechanism requires recursive computation across the citation network and, even in its more tractable approximate form, is ``particularly unkind to junior researchers''~\cite{Hirsch2010}.
An alternative that inherits the simplicity of the h-index is to allocate citations fractionally among authors.

Derek de Solla Price~\cite{Price1981} advocated distributing credit for a scientific publication among all authors to preclude undesirable publication practices: ``The payoff in brownie points of publications or citations must be divided among all authors listed on the byline, and in the absence of evidence to the contrary it must be divided equally among them. [...] If this is strictly enforced it can act perhaps as a deterrent to the otherwise pernicious practice of coining false brownie points by awarding each author full credit for the whole thing.''~\cite{Price1981}.
Since the introduction and broad adoption of the h-index~\cite{Hirsch2005}, many variants and related measures have been proposed~\cite{PanaretosMalesios2009,Waltman2016,Abambres2016}. Some of these implement fractional allocation.
Batista et al.~\cite{Batista2006} present a normalization of the h-index by the average number of authors of papers in the h-core.
Wan et al.~\cite{Wan2007} perform a similar normalization, but use the square root of the average authors of papers in the h-core.
Chai et al.~\cite{Chai2008} describe a variant of the h-index that is based on citation counts normalized by the square root of the number of authors per paper.
Egghe~\cite{Egghe2008} introduces alternative versions of the h- and g-index (see supplementary information) that use citation counts normalized by the number of authors. Egghe's version of the h-index corresponds to the h-frac measure that we find to be particularly effective in our experiments. Note that the work of Egghe is purely theoretical and does not include any experiments with real bibliographic data~\cite{Egghe2008}.
Schreiber~\cite{Schreiber2008,Schreiber2008a} presents an alternative fractional allocation measure. Instead of using normalized citation counts, Schreiber proposes to first compute alternative (``effective'') publication ranks that are divided by the number of authors.
These effective ranks are then used to determine the $\textrm{h}_\textrm{m}$-index, akin to computing the h-index with unmodified publications ranks. A related alternative has also been proposed for the g-index~\cite{Schreiber2009,Schreiber2010}.
Other variants that apply different fractional allocation schemes can also be found in the literature~\cite{Prathap2011,Tol2011,Galam2011,Rousseau2014}.
While there exist bibliometric tools that implement fractional versions of the h-index~\cite{Harzing2007,Kozlowski2019}, we are not aware of published systematic empirical evaluation of fractional allocation measures with real bibliographic data, on a large scale (millions of articles), and across multiple scientific fields and data platforms. We contribute such an evaluation.
Among other measures, we experimentally evaluate h-frac alongside the scientometric measures of Batista et al.~\cite{Batista2006} ($\textrm{h}_\textrm{I}$), Schreiber~\cite{Schreiber2008,Schreiber2008a} ($\textrm{h}_\textrm{m}$), Wan et al.~\cite{Wan2007} ($\textrm{h}_\textrm{p}$), and Chai et al.~\cite{Chai2008} ($\textrm{h}_\textrm{ap}$).

\begin{figure*}[t!]
	\centering
	\begin{subfigure}[t]{0.375\textwidth}
		\centering
		\caption{}
		\label{fig:effect_and_predict_summary}
		\includegraphics[width=\textwidth]{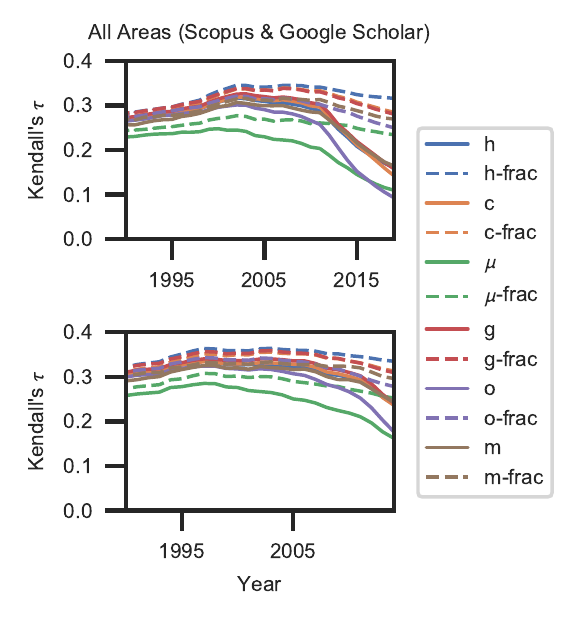}
	\end{subfigure}%
	\begin{subfigure}[t]{0.375\textwidth}
		\centering
		\caption{}
		\label{fig:effect_and_predict_other}
		\includegraphics[width=\textwidth]{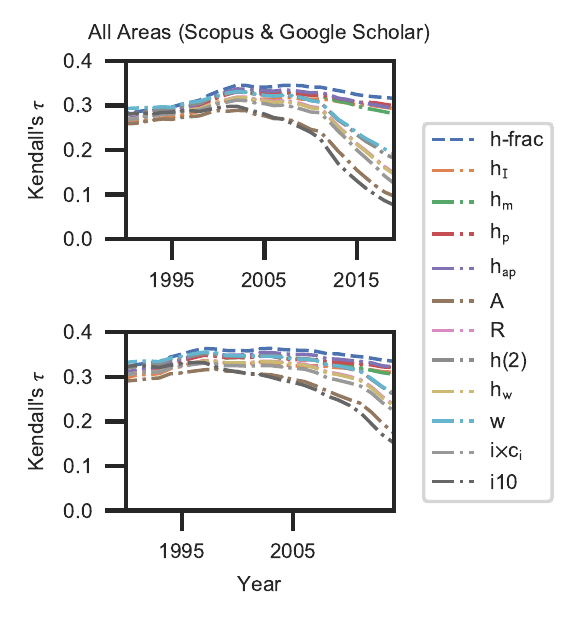}
	\end{subfigure}
	\caption{Effectiveness and predictive power of scientometric measures.
		In each subfigure, the top row depicts the correlation of bibliometric indicators and scientific awards, and the bottom row shows the predictive power five years into the future.
		\sflabel{a}~Evaluation across all research areas and data platforms (Scopus and Google Scholar).
		\sflabel{b}~Evaluation of h-frac alongside additional measures across all research areas and data platforms.
		}
	\label{fig:effect_and_predict}
\end{figure*}

Fig.~\ref{fig:effect_and_predict_summary}(top) contrasts the effectiveness of fractional allocation measures and traditional ones across all research fields and data platforms. We again measure the correlation of rankings induced by different bibliometric measures and scientific reputation as evidenced by awards bestowed by the scientific community.
Detailed results for the individual research areas can be found in Supplementary Fig.~\ref{fig:criteria_all}(left).

We find that fractional measures are significantly more effective correlates of scientific awards than unnormalized indicators such as the h-index.
The fractional analogue of the h-index, h-frac, is the most effective measure across datasets (average ${\tau = 0.32}$ in 2019, compared to $0.16$ for the h-index; see Supplementary Table~\ref{tab:effectiveness}(top)). The effectiveness of fractional allocation measures is more stable over time than the effectiveness of their traditional counterparts. (For h-frac, average $\tau=0.28$ in 1989 and $0.32$ in 2019; for the h-index, average $\tau=0.27$ in 1989 and $0.16$ in 2019.)

\subsection*{Predictive power and other measures}

Next we evaluate the predictive power of different bibliometric measures. Prior studies have largely focused on the ability of measures to predict their own future values, or those of other bibliometric indicators~\cite{Hirsch2007,Acuna2012,Penner2013}. In contrast, we study the ability of an indicator to predict a scientist's future reputation as evidenced by scientific awards. (Hirsch recognized this as a meaningful goal when he wrote ``how likely is each candidate to become a member of the National Academy of Sciences 20 years down the line?'', but did not operationalize this~\cite{Hirsch2007}.) We measure the correlation of rankings induced by scientometric indicators in a given year (e.g.\ 2010) with rankings induced by awards in a future year (e.g.\ 2015). Higher correlation implies stronger ability to predict future scientific reputation based on present-day bibliometric data.

Fig.~\ref{fig:effect_and_predict_summary}(bottom) reports predictive power five years into the future. The results are summarized across all research fields and data sources.
The predictive power of the h-index has declined since its introduction (average $\tau= 0.32$ in 2004 versus $0.24$ in 2014). Other traditional indicators have also declined in effectiveness. Fractional measures are more predictive. h-frac has the highest predictive power across datasets and its predictive power is stable over time (average $\tau$ is $0.34$ in 1994, $0.36$ in 2004, and $0.33$ in 2014).

We further evaluate h-frac alongside an extensive list of other scientometric measures~\cite{PanaretosMalesios2009,Batista2006,Schreiber2008,Schreiber2008a,Wan2007,Chai2008,Jin2006,Jin2007,Kosmulski2006,Egghe2008a,Wu2009,Kosmulski2007,scholar}.
The results are summarized in Fig.~\ref{fig:effect_and_predict_other}.
Measures that integrate some form of normalization by the number of coauthors (h-frac, $\text{h}_\text{I}$, $\text{h}_\text{m}$, $\text{h}_\text{p}$, $\text{h}_\text{ap}$) outperform measures that do not apply such normalization.
h-frac is the best-performing measure in terms of both correlation with scientific awards and predictive power.

\subsection*{Robustness of the findings}

We now test the robustness of the findings in a number of additional controlled experiments.

\begin{figure*}[t!]
	\begin{minipage}{0.25\textwidth}
		\begin{subfigure}[t]{\textwidth}
			\centering
			\caption{}
			\label{fig:analysis_ref}
			\includegraphics[width=\textwidth]{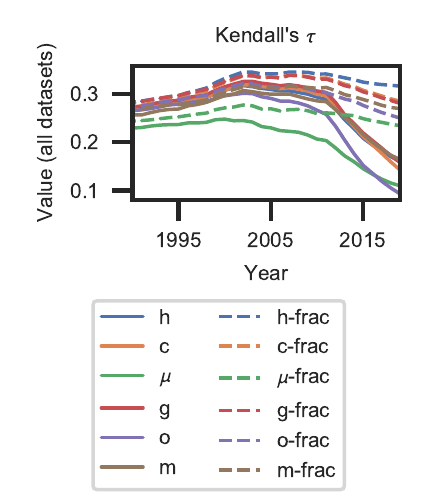}
		\end{subfigure}
	\end{minipage}
	\begin{minipage}{0.75\textwidth}
	\begin{subfigure}[t]{\textwidth}
		\centering
		\caption{}
		\label{fig:analysis_criteria}
		\includegraphics[width=\textwidth]{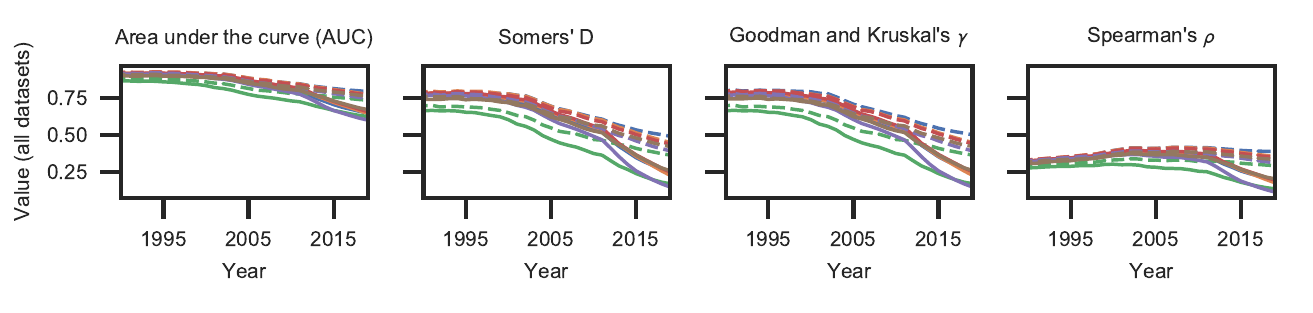}
	\end{subfigure}\\[-6mm]
	\begin{subfigure}[t]{\textwidth}
		\centering
		\caption{}
		\label{fig:analysis_awards}
		\includegraphics[width=\textwidth]{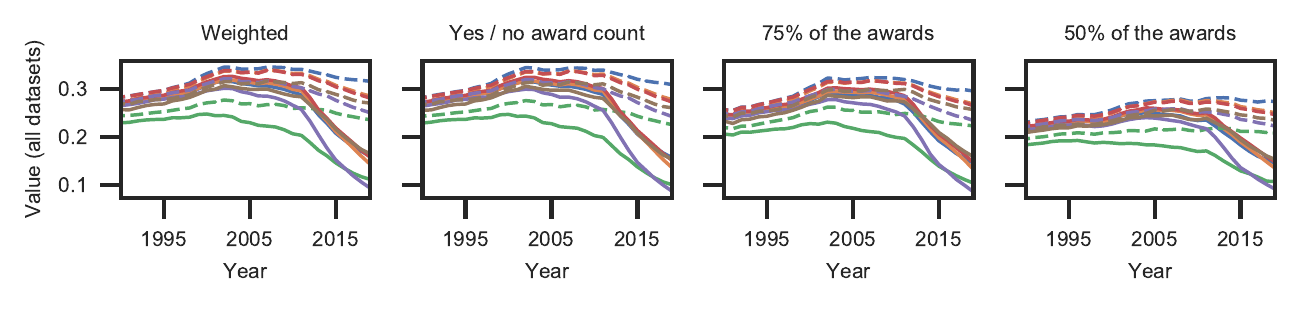}
	\end{subfigure}\\[-6mm]
	\begin{subfigure}[t]{\textwidth}
		\centering
		\caption{}
		\label{fig:analysis_subsets}
		\includegraphics[width=\textwidth]{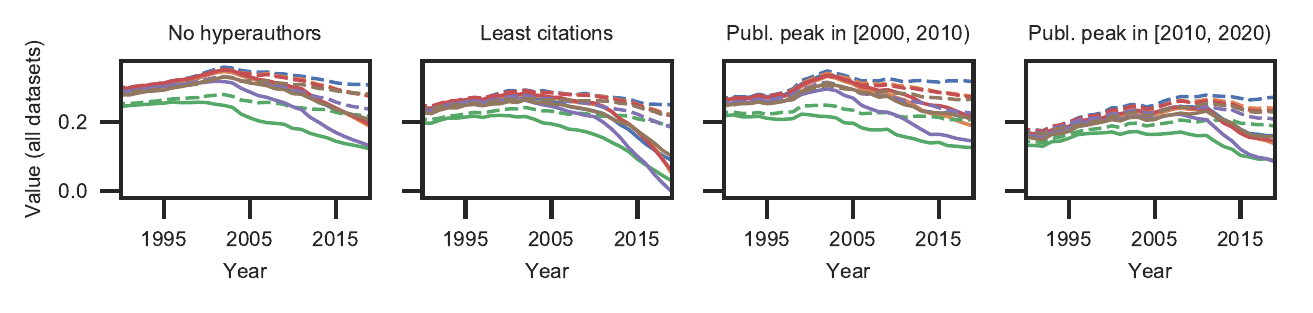}
	\end{subfigure}
	\end{minipage}
	\caption{Controlled experiments that test the robustness of the findings.
	\sflabel{a}~Reference result from the main experiments (cf.\ Fig.~\ref{fig:effect_and_predict_summary}(top)).
	\sflabel{b}~Corresponding results with other correlation statistics.
		(\textbf{c} and \textbf{d})~Results in different conditions: using subsets of awards, researchers, and different mechanisms for counting awards.
	}
	\label{fig:robustness}
\end{figure*}

First, we repeat the experiments with different correlation statistics (see supplementary information). The results are summarized in Fig.~\ref{fig:analysis_criteria}, and detailed results for all research areas and data platforms can be found in the supplementary materials (Supplementary Fig.~\ref{fig:criteria_all}).
Fractional measures continue to outperform their traditional counterparts, and h-frac is the most reliable indicator.

Next we analyze robustness with respect to the set of scientific awards considered in our datasets.
Our main experiments treated all awards equally, and ranked scientists by the total number of awards received. For example, a Nobel prize was given the same weight as membership in the National Academy of Sciences, and a scientist with two awards was ranked higher than a scientist with one award.
To examine whether our findings are sensitive to this choice, we repeat the experiments under different conditions.
First, we assign 10 times higher weight to awards with 100 or fewer laureates. (See Supplementary Table~\ref{tab:awards}.)
Second, we evaluate a design in which the number of awards does not affect a scientist's ranking: a scientist with an award of any kind is ranked higher than a scientist with no awards, but all scientists with one or more awards are ranked equally. The results are summarized in Fig.~\ref{fig:analysis_awards}(left) and presented in detail in the supplementary materials (Supplementary Fig.~\ref{fig:awards_all}). Our findings hold for both conditions. (The results remain consistent for other weighting factors and thresholds as well.)

To further assess sensitivity, we repeat the experiments with random subsets of awards (using 75$\%$ and 50$\%$ of awards in our database). The results are reported in Fig.~\ref{fig:analysis_awards}(right) and Supplementary Fig.~\ref{fig:awards_all}. Our findings again hold.
This demonstrates the robustness of our findings with respect to the considered awards and the matching procedure. (See supplementary information.)

Is the decline in the effectiveness of the h-index and other traditional scientometric measures solely due to the rise of hyperauthorship?
To investigate this hypothesis, we curtail the effect of hyperauthorship by reproducing the experiments with the set of authors who have at most 100 coauthors per paper on average. The results in Fig.~\ref{fig:analysis_subsets}(left) show that our findings hold in this condition as well: we see a strong decline in the effectiveness of traditional measures, in contrast to the stable performance of their fractional counterparts. Hyperauthors appear to be an extreme manifestation of a broader shift in publication patterns. Hyperauthors themselves are not the main cause of the decline in the effectiveness of the h-index and other measures, and pruning hyperauthors from datasets does not avert this decline.

Next we perform experiments with different subsets of researchers.
First we remove the most highly-cited researchers in our datasets and repeat the experiments with the bottom 50$\%$ of researchers in each field by number of citations.
This examines whether our findings hold for researchers that are not at the very top of their fields in terms of citations.
Then we analyze the effect of the main time period of a scientist's work. (Details on the temporal coverage of the authors in our dataset can be found in Supplementary Fig.~\ref{fig:data_cum}.)
To this end, we choose subsets of researchers that are active at different periods of time.
Specifically, we test the subset of researchers whose peak productivity (in terms of number of publications) occurs during the years $\left[2000, 2010\right)$, and another subset whose peak productivity occurs during the years $\left[2010, 2020\right)$.

The results are summarized in Fig.~\ref{fig:analysis_subsets} and given in detail in Supplementary Fig.~\ref{fig:subsets_all}. Our main findings are robust to all these perturbations and hold in all conditions: fractional allocation measures always outperform their traditional counterparts, and h-frac is the most reliable bibliometric indicator across all conditions.

\subsection*{Correlation between scientometric measures}

\begin{figure*}[t!]
	\centering
	\begin{subfigure}[c]{0.6\textwidth}
		\centering
		\caption{}%
		\label{fig:metric_correlation}
		\includegraphics[width=\textwidth]{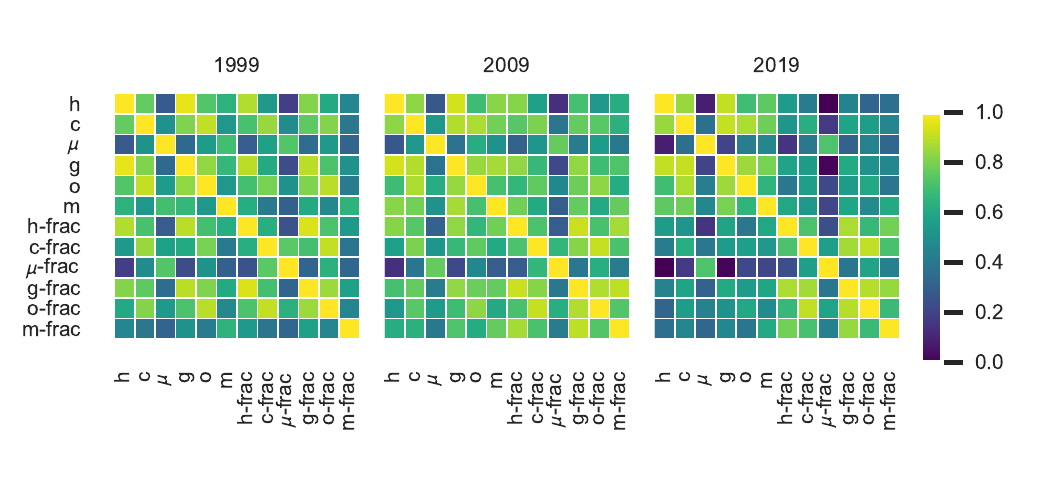}
	\end{subfigure}%
	\begin{subfigure}[c]{0.4\textwidth}
		\centering
		\caption{}
		\label{fig:metric_correlation_yearly}
		\includegraphics[width=\textwidth]{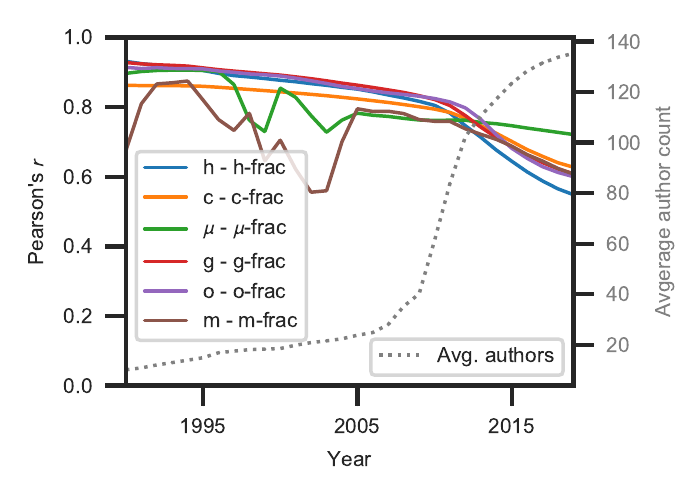}
	\end{subfigure}
	\caption{Correlation between scientometric measures.
		\sflabel{a}~Correlation matrices of scientometric measures in the years 1999, 2009, 2019.
		\sflabel{b}~Temporal evolution of correlations between traditional measures and their fractional counterparts.
	}
	\label{fig:correlation}
\end{figure*}

Our experiments indicate that fractional allocation measures are superior to their traditional counterparts. To analyze this further, we investigate the correlation between different scientometric measures~\cite{RadicchiCastellano2013,Ioannidis2016}. To this end, we compute the correlation between each pair of measures, aggregated over all datasets (Fig.~\ref{fig:metric_correlation}). To interpret the results, we consider three different 6x6 blocks in the correlation matrices:

\begin{enumerate}[label=(\roman*)]
  \item
  The \emph{lower right} block summarizes the correlations between the fractional measures. It is quite stable over the years. All fractional measures are moderately correlated, with the exception of $\mu$-frac. The lower correlation of $\mu$-frac with the other fractional measures can be explained by the explicit normalization by the number of publications in $\mu$-frac, which is absent in the other measures. As can be seen in the preceding results, $\mu$-frac is the worst-performing measure among the fractional ones.
	\item
  The \emph{upper left} block summarizes the correlations between the traditional measures. These correlations are stable over time. The traditional measures are moderately correlated with each other, again with the exception of $\mu$. This can again be attributed to the explicit normalization by the number of publications in $\mu$.
	\item
  The \emph{lower left} block captures the correlations between the traditional and fractional measures. Notably, we observe that these correlations decrease significantly from 2009 to 2019. All correlation values decrease, including the correlations between the traditional measures and their direct fractional counterparts (the diagonal in the lower-left block). The measures $\mu$ and $\mu$-frac stand out again, which can be attributed to the same factors as in the other blocks.
\end{enumerate}

Why have the traditional and fractional measures become less correlated over time? We examine the temporal evolution of correlations between traditional measures and their fractional counterparts at finer granularity (Fig.~\ref{fig:metric_correlation_yearly}). We see that the correlation decreases over time, with accelerated decline after 2010. Concurrently, the average number of authors per publication rises significantly. The two trends are strongly correlated. Since accounting for the number of authors per publication is the central feature that distinguishes fractional measures from their traditional counterparts, we attribute the diminishing correlation between the measures to the changing publication culture, as reflected in the dramatic increase in the average number of authors per paper.

\subsection*{Further analysis}
\begin{figure*}[t!]
	\centering
	\begin{subfigure}[t]{\textwidth}
		\centering
		\caption{}
		\label{fig:ind_scopus_physics}
		\includegraphics[width=\textwidth]{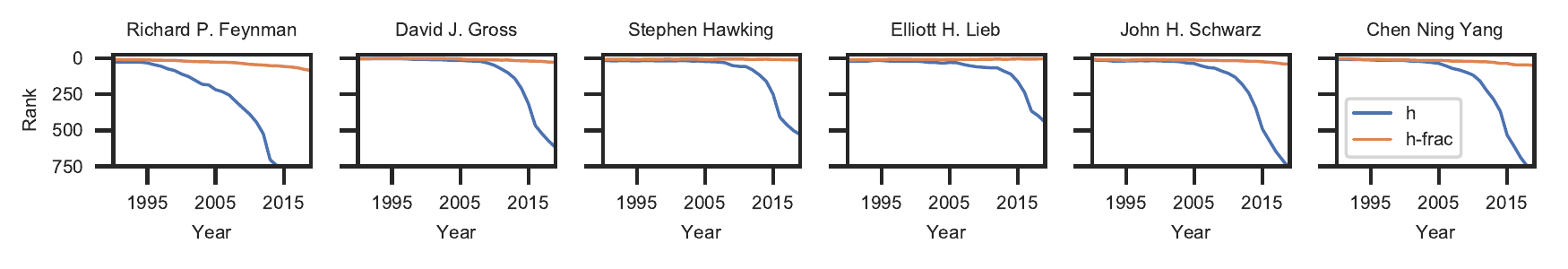}
	\end{subfigure}\\[-5mm]
	\begin{subfigure}[t]{\textwidth}
		\centering
		\caption{}
		\label{fig:annotated_scatter_scopus_physics}
		\includegraphics[width=\textwidth]{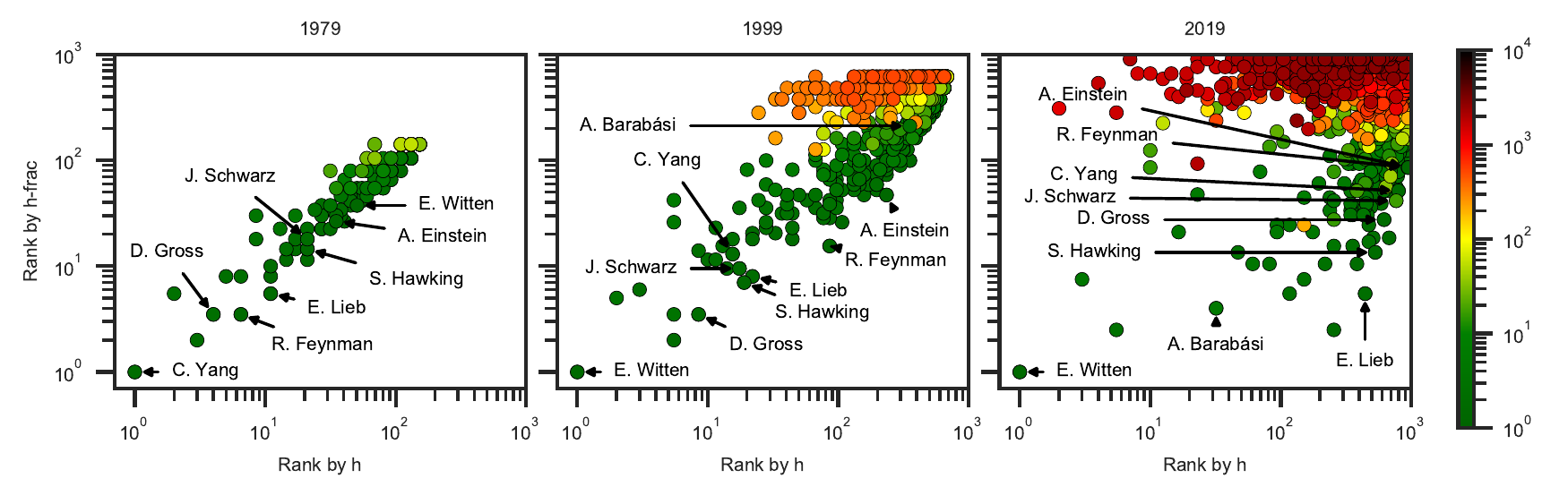}
	\end{subfigure}
	\centering
	\begin{subfigure}[t]{0.45\textwidth}
		\centering
		\caption{}
		\label{fig:sankey_h}
		\includegraphics[width=\textwidth]{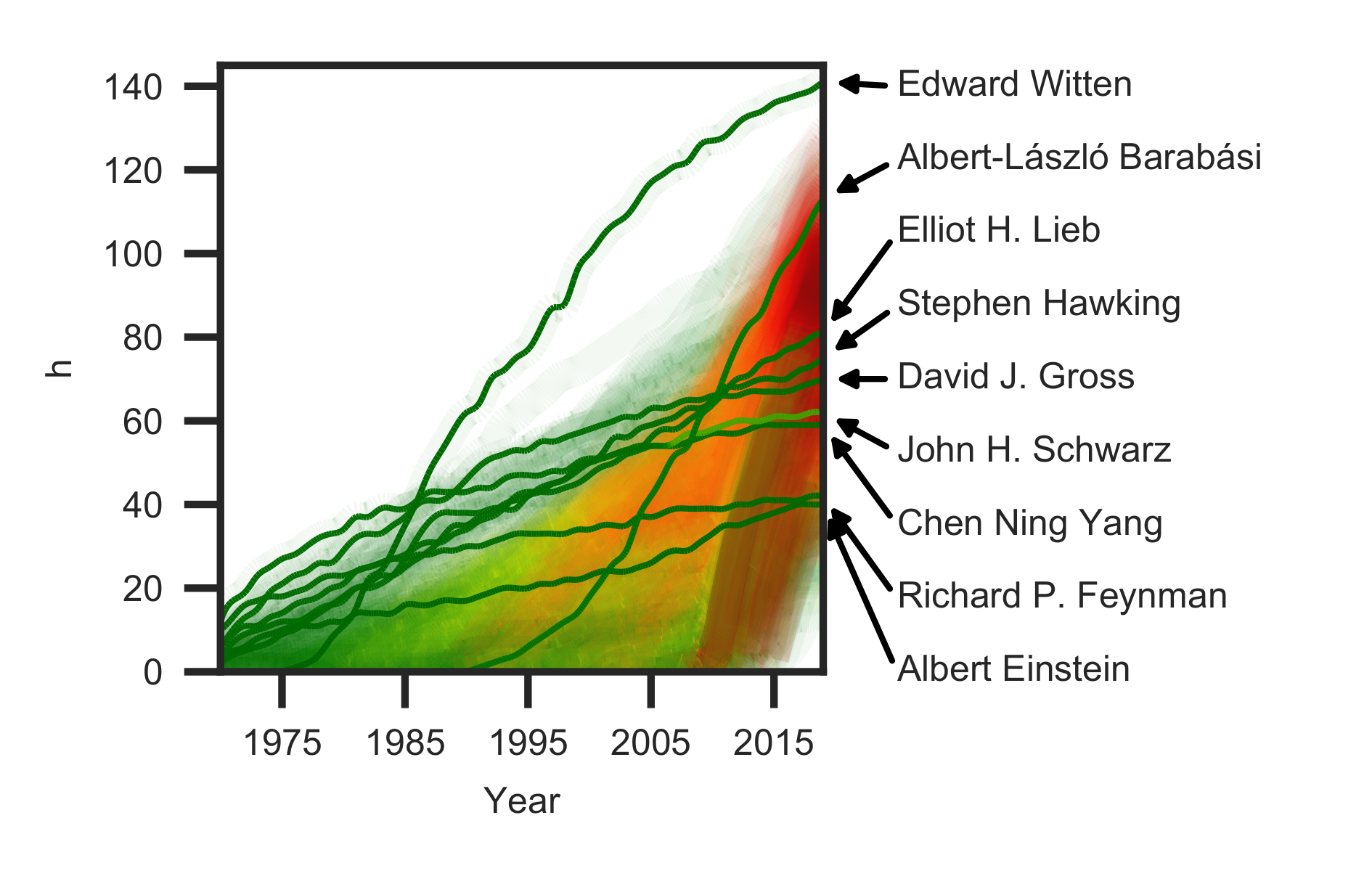}
	\end{subfigure}%
	\begin{subfigure}[t]{0.45\textwidth}
		\centering
		\caption{}
		\label{fig:sankey_h_frac}
		\includegraphics[width=\textwidth]{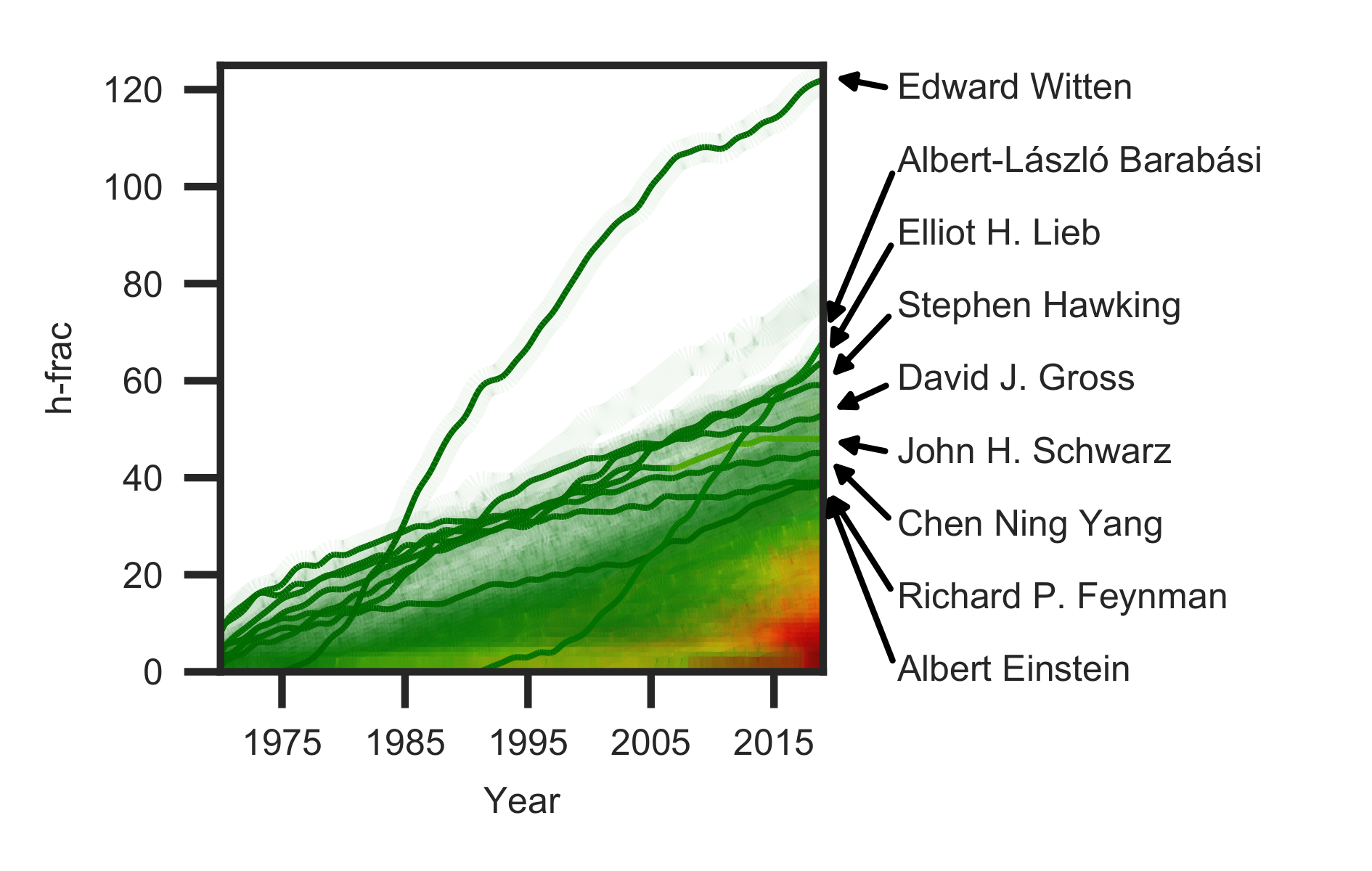}
	\end{subfigure}%
	\begin{subfigure}[t]{0.1\textwidth}
		\centering
		\caption*{}
		\includegraphics[width=\textwidth]{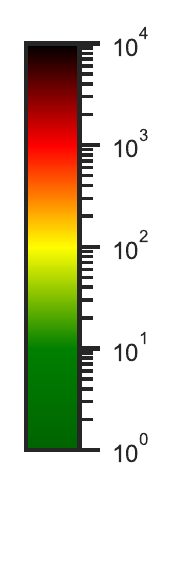}
	\end{subfigure}\\[-5mm]
	\begin{subfigure}[t]{0.5\textwidth}
		\centering
		\caption{}
		\label{fig:dist_h_frac_scopus}
		\includegraphics[width=\textwidth]{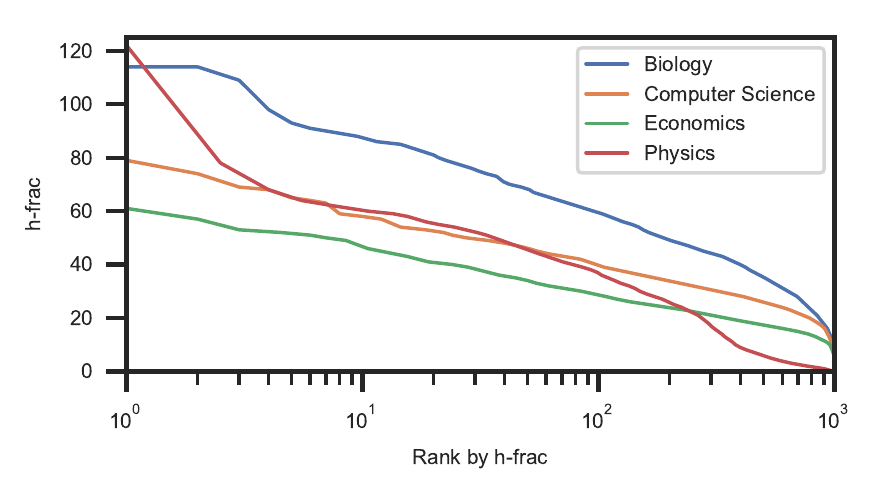}
	\end{subfigure}%
	\begin{subfigure}[t]{0.5\textwidth}
		\centering
		\caption{}
		\label{fig:num_authors_top_physicists}
		\includegraphics[width=\textwidth]{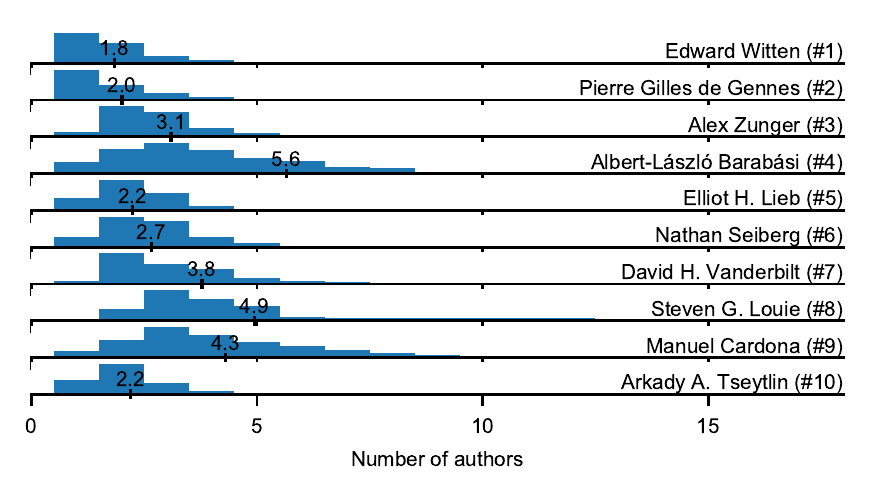}
	\end{subfigure}
	\caption{Further analysis.
	\sflabel{a}~Ranking induced by h and h-frac for a number of scientists in the Scopus physics dataset.
	\sflabel{b}~Comparison of rankings induced by h and h-frac in the Scopus physics dataset. Scientists are color-coded by the average number of coauthors per publication.
	\sflabel{c}~Evolution of the h-index of each scientist in the Scopus physics dataset over time. Each scientist is a curve. Color represents the average number of coauthors per publication.
	\sflabel{d}~Evolution of h-frac over time.
	\sflabel{e}~Distribution of h-frac values in each field of research.
	\sflabel{f}~Distribution of the number of authors per publication for 10 physicists with the highest h-frac in 2019.
	}
	\label{fig:further_analysis}
\end{figure*}

Fig.~\ref{fig:ind_scopus_physics} provides a number of case studies that highlight the stability of h-frac and the deterioration of the h-index over time. These case studies are further illustrated in Fig.~\ref{fig:annotated_scatter_scopus_physics}.
The evolution of h and h-frac values over time is visualized in Figs.~\ref{fig:sankey_h} and~\ref{fig:sankey_h_frac}. Hyperauthors (red) acquire increasingly high h-indices over time, commonly rising above 80 by 2019. In contrast, their h-frac values remain low, predominantly less than 20. Fig.~\ref{fig:dist_h_frac_scopus} visualizes the distribution of h-frac values in the four fields of research. The top 100 scientists have h-frac values of 59 and higher in biology, 39 and higher in computer science, 37 and higher in physics, and 29 and higher in economics.

Fig.~\ref{fig:num_authors_top_physicists} examines in detail the output of the 10 physicists with the highest h-frac in 2019. The data suggests that the h-frac measure is not antithetical to collaboration, which is associated with scientific progress~\cite{Wuchty2007,Dong2017,Wu2019}. Among physicists with the highest h-frac are prolific collaborators such as Albert-L\'aszl\'o Barab\'asi (\#4, 5.6 authors per publication on average), Steven G. Louie (\#8, 4.9 authors per publication on average), and Manuel Cardona (\#9, 4.3 authors per publication on average).

\section*{Discussion}

We have conducted a large-scale systematic analysis of scientometric measures.
We have demonstrated that commonly used measures of a scientist's impact have become less effective as correlates and predictors of scientific reputation as evidenced by scientific awards. The decline in the effectiveness of these measures is associated with changing authorship patterns in the scientific community, including the rise of hyperauthorship.
We have also demonstrated that fractional allocation of citations among coauthors improves the robustness of scientometric measures. In particular, the h-frac, a fractional analogue of the h-index, is the most reliable measure across different experimental conditions.

Our analysis did not uncover unreasonable penalization of collaboration among researchers by fractional allocation measures. Fractional allocation does make explicit the expectation that each author makes a meaningful contribution to the publication's impact.
In the words of Derek de Solla Price, ``Those not sharing the work, support, and responsibility do not deserve their names on the paper, even if they are the great Lord Director of the Laboratory or a titular signatory on the project. Any time you take a collaborator you must give up a share of the outcome, and you diminish your own share. That is as it should be; to do otherwise is a very cheap way of increasing apparent productivity.''~\cite{Price1981}.
Our study indicates that fractional allocation neutralizes the inflationary effects of hyperauthorship on bibliometric impact indicators, but continues to reward collaborative production of impactful scientific research~\cite{Wuchty2007,Dong2017,Wu2019}.

A number of aspects of bibliometric impact indicators have not been addressed in our study. One is the normalization of bibliometric indicators across different fields, so as to enable direct comparison of scientists across fields with different publication and citation patterns~\cite{Waltman2015,Waltman2016}. Another is the presence of self-citations and whether such citations should be handled differently~\cite{Schreiber2009a,Waltman2016}. Likewise we have not addressed the role of author order and whether this order should be taken into account in automatically allocating credit for a publication's impact~\cite{Marusic2011,Waltman2016}. These are interesting avenues for future work.

Our work has both near-term and long-term implications. In the near term, our work indicates that the use of the h-index in assessing individual scientific impact should be reconsidered, and that h-frac can serve as a more robust alternative. This can ameliorate distortions introduced by contemporary authorship practices, lead to a more effective allocation of resources, and facilitate scientific discovery. In the longer term, our data, methodology, and findings can inform the science of science~\cite{Fortunato2018,Ioannidis2015} and support further quantitative analysis of research, publication, and scientific accomplishment.

An interactive visualization of our work can be found at \url{https://h-frac.org}.

    \section*{Methods}

\subsection*{Highly-cited researchers}
We construct a dataset of highly-cited researchers in four research fields: biology, computer science, economics, and physics.
To begin, we retrieve a set of highly-cited researchers in each field via Google Scholar. To this end, we query Google Scholar with labels that are characteristic of different research areas (Supplementary Fig.~\ref{fig:scholar_search_labels}). The retrieved authors are sorted by the number of citations: most highly cited researchers appear first. However, the results are noisy because the queries retrieve all authors that feature the queried keyword phrases in their profiles. For example, a physicist who features ``high performance computing'' as a keyword phrase in their profile would be retrieved by the corresponding query. Since ``high performance computing'' is one of our queries for computer science researchers, the physicist would, in the absence of further validation, be added to the computer science dataset.

To clean up the initial lists compiled via Google Scholar, we cross-reference them with the Scopus database. A scientist's Scopus profile indicates their primary research area. We use this primary research area to filter the initial lists. To this end, we need to match author profiles in Google Scholar with Scopus profiles.
To perform the association, we first create a set of candidate matches by querying the Scopus database with the researcher's name. To obtain the query name, we clean the Google Scholar profile name via simple heuristics (e.g.\ remove extraneous information such as links or affiliation names).
To reduce false positives, we limit the candidates to Scopus profiles with more than 50 papers (more than 30 papers for economics). To perform the actual matching, we analyze the top 100 papers (sorted by citation counts) of the different candidate profiles. If we find at least three matching paper titles in the Scholar and Scopus profiles, we associate the two profiles.

After matching, we filter the authors in each field by their primary subject area in Scopus (Supplementary Fig.~\ref{fig:search_subject_areas}).
After filtering, we retain the top 1,000 authors in each field. This filtered set is derived from the top 1,186 Google Scholar profiles in biology, 1,711 in computer science, 1,632 in economics, and 1,296 in physics. This means that, in aggregate, more than two thirds of the initial Google Scholar profiles are matched to corresponding Scopus profiles with the desired primary subject area. Authors that could not be matched or do not have the requisite primary subject area are removed from the corresponding list. (They may still be retained in a list for a different field; e.g.\ physics rather than computer science.) One attribute of our filtering procedure is that the lists of authors in the four fields are disjoint: a scientist is only included in at most one list.

\subsection*{Google Scholar data}
For all 4,000 researchers, we collect their Google Scholar publications including citation data~\cite{scholar}.
In particular, we collect (for each publication) the publication year, the number of authors, and the number of citations per year. We filter out certain publications: (i)~publications that do not list authors or the publication year, (ii)~patents, and (iii)~duplicates marked by Google Scholar.
Moreover, we noticed that the publication date and the citation years in Google Scholar are sometimes inconsistent: a publication is sometimes cited \emph{before} is was published. As a remedy, we take the minimum of the publication year and the year of the first citation as the effective publication year.

We also noticed that Google Scholar generally under-reports the number of authors for publications with large author sets. Manual inspection indicates that Scholar does not record all authors, but only the first $\sim$150 authors. In particular, the maximal value of the average author count in the Scholar dataset is 230, versus 3,130 in Scopus.
This is an important limitation of the Scholar data that has to be kept in mind. The consistency of our findings across the Scholar and Scopus datasets, in spite of the truncated author counts in the Scholar data, indicates that our findings are robust to such noise and bias in the data.

\subsection*{Scopus data}
Similar to the Google Scholar data, we collect for each of the 4,000 authors their Scopus publications with citation data~\cite{scopus}.
Since the Scopus data is significantly less noisy than the Scholar data, no special data cleaning and filtering are required.

\medskip
One salient difference between the datasets is that the Google Scholar datasets contain approximately twice as many publications and citations than the Scopus datasets. One contributing factor is that Scopus indexes only a subset of the venues crawled by Google Scholar. For example, Scopus does not index online repositories such as arXiv. In agreement with prior studies, we have found Google Scholar data to be both broader and noisier than Scopus~\cite{Waltman2016}. The consistency of our findings across the Scholar and Scopus datasets highlights their robustness.

\subsection*{Award data}
We use awards bestowed by the scientific community as indicators of scientific reputation. To this end, we consider highly selective distinctions, some of which span multiple scientific fields, such as membership in the National Academy of Sciences, and some of which are field-specific, such as fellowship of the Econometric Society (Supplementary Fig.~\ref{fig:winner_all}, Supplementary Table~\ref{tab:awards}, and \url{https://h-frac.org/dataset-s1}).

Our award data collection procedure begins by compiling complete lists of laureates for each award from the respective web sites. (This is nontrivial since it requires customized parsing techniques for each award.)
Next, we search these lists of laureates for names in our datasets. This search is based on the surname and the initials from each Scopus author profile in our dataset. This yields a list of candidate matches. We then manually check all candidate matches, considering the author details in the Scopus profile, such as name variations, affiliations, and subject areas, as well as details extracted from the corresponding award pages, such as bio, affiliation, and country. (Supplementary Figs.~\ref{fig:awards}, A, B, and Supplementary Table~\ref{tab:awards}).

For each laureate, we also retain the year in which the award was conferred. This is central to our measurement of correlation and predictive power over time.

    \printbibliography[title=References,segment=\therefsegment]

\end{refsegment}

\section*{Author contributions statement}

V.K. conceived and directed the project. D.H. collected the data, implemented the methods, and performed experiments. V.K. and D.H. designed experiments, analyzed data, and wrote the paper.

\section*{Additional information}

The authors declare no competing interests.

\begin{refsegment}


\renewcommand{\thefigure}{S\arabic{figure}}
\renewcommand{\thetable}{S\arabic{table}}

\setcounter{figure}{0}    
\setcounter{table}{0}    

\section*{Supplementary Information}

\subsection*{Data Collection}

\subsubsection*{Scholar Data}
In total, we collect 2,624,994 (valid) publications in Google Scholar that are collectively cited 220,783,854 times.
The distribution of the publications and citations among the research fields is as follows (Fig.~\ref{fig:data_cum}(bottom)): biology accounts for 31\% of the publications and 38\% of the citations, computer science for 22\% and 17\%, economics for 12\% and 10\%, and physics for 35\% and 35\%.
Our dataset offers yearly granularity from 1970 onwards.

\subsection*{Scopus Data}
\begin{figure}
    \centering
	\includegraphics[width=\textwidth]{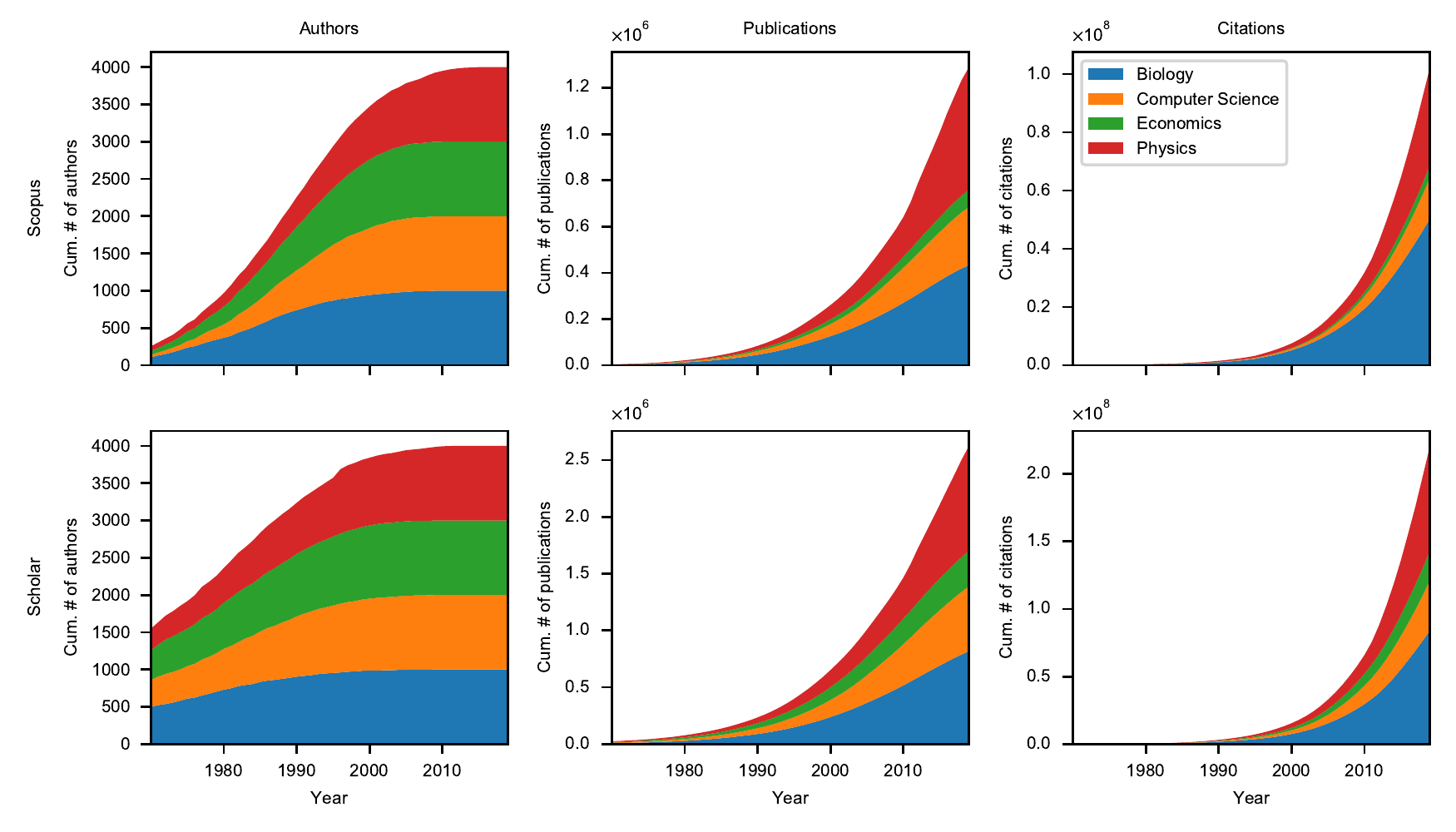}
	\caption{Overview of Scopus and Google Scholar datasets.
		 Scholar (top) and Google Scholar (bottom) datasets. From left to right: Cumulative number of authors, publications, and citations per year, from 1970 onwards. Authors are considered present in the database if they have at least one publication recorded by the considered year.
	}
	\label{fig:data_cum}
\end{figure}

In total, the Scopus dataset comprises 1,290,219 publications with 102,405,086 citations. The distribution of publications and citations is as follows (Fig.~\ref{fig:data_cum}(top)): biology accounts for 34\% of publications and 49\% of citations, computer science for 20\% and 14\%, economics for 6\% and 4\%, and physics for 41\% and 33\%.

\subsection*{Award Data}
\begin{figure}
	\centering
	\begin{subfigure}[c]{0.66\textwidth}
		\centering
		\caption{}
		\label{fig:winner_all}
		\includegraphics[width=\textwidth]{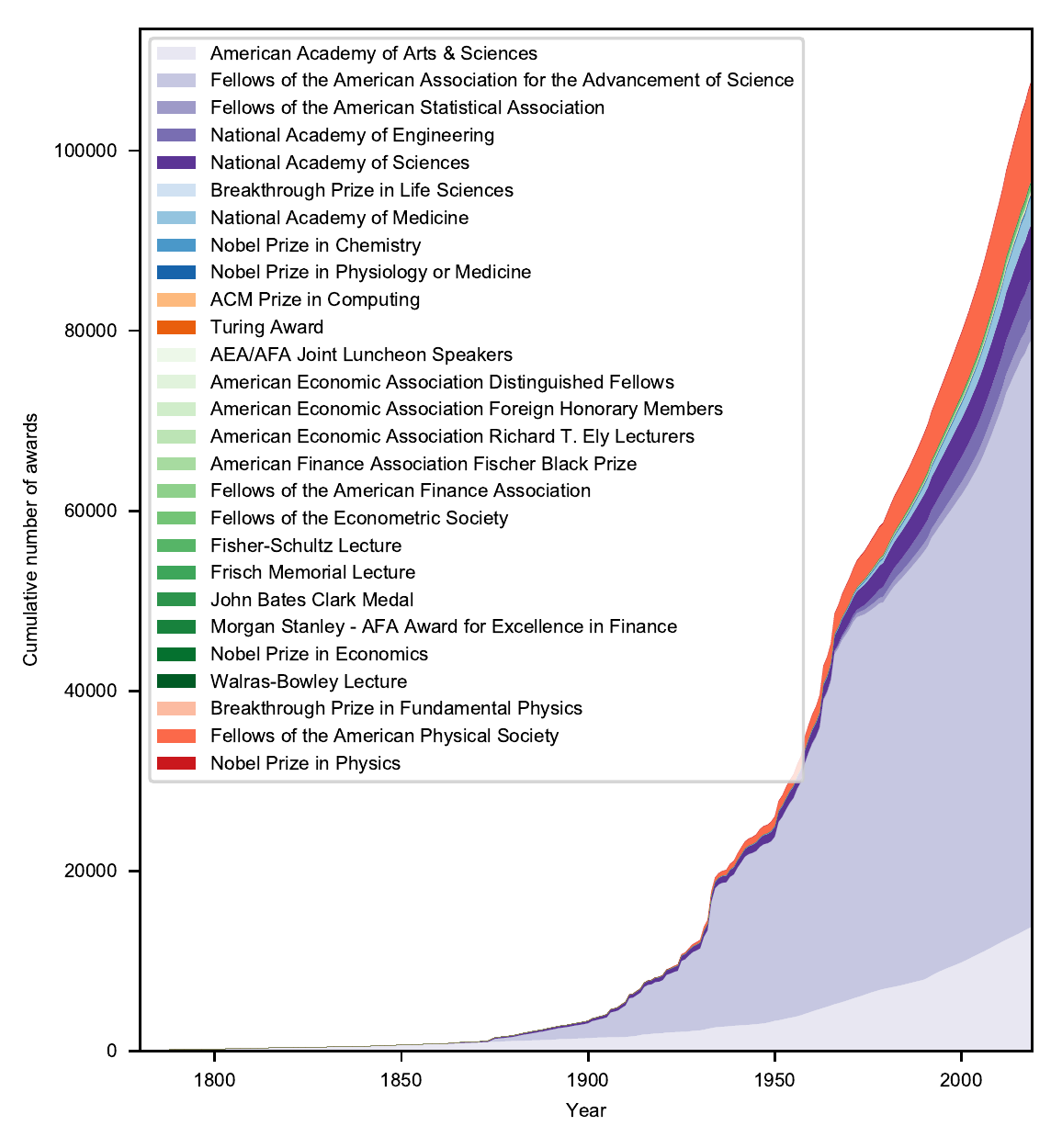}
	\end{subfigure}
	\begin{subfigure}[c]{0.33\textwidth}
		\centering
		\caption{}
		\label{fig:awards_all}
		\includegraphics[width=\textwidth]{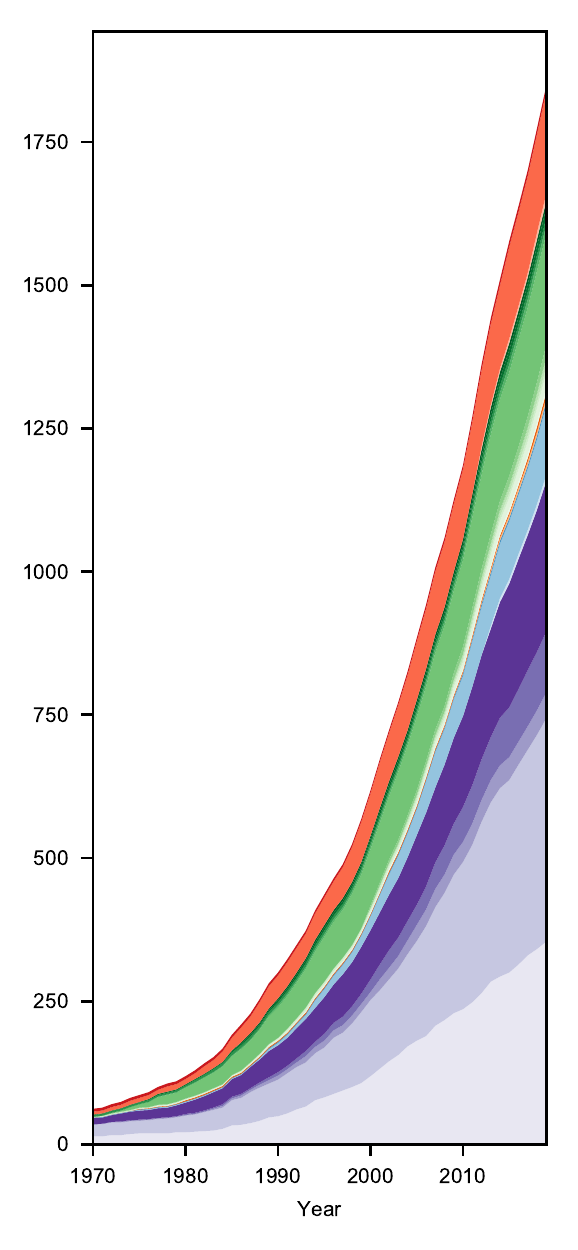}
	\end{subfigure}
	\begin{subfigure}[t]{0.495\textwidth}
		\centering
		\caption{}
		\label{fig:awards_yearly}
		\includegraphics[width=\textwidth]{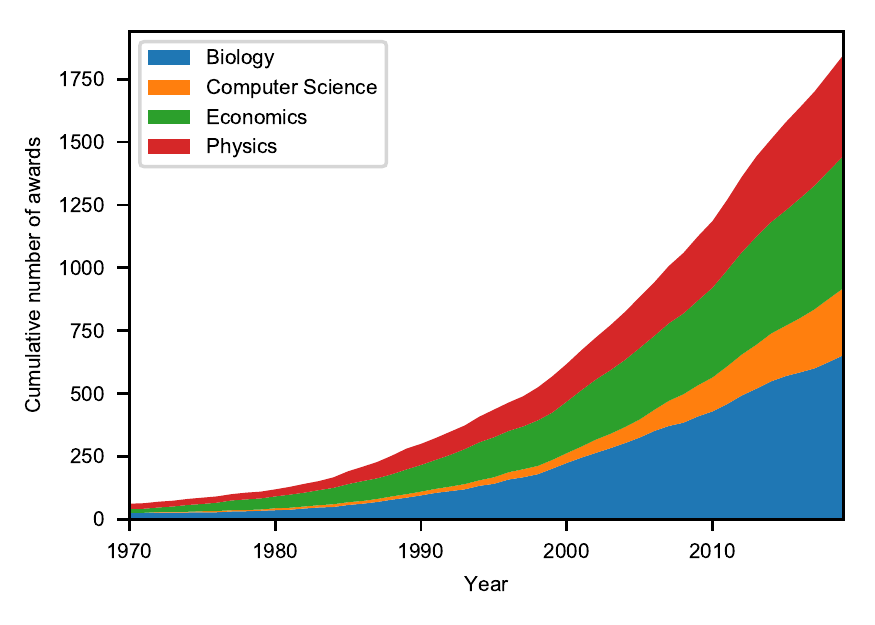}
	\end{subfigure}
	\begin{subfigure}[t]{0.495\textwidth}
		\centering
		\caption{}
		\label{fig:awards_dist}
		\includegraphics[width=\textwidth]{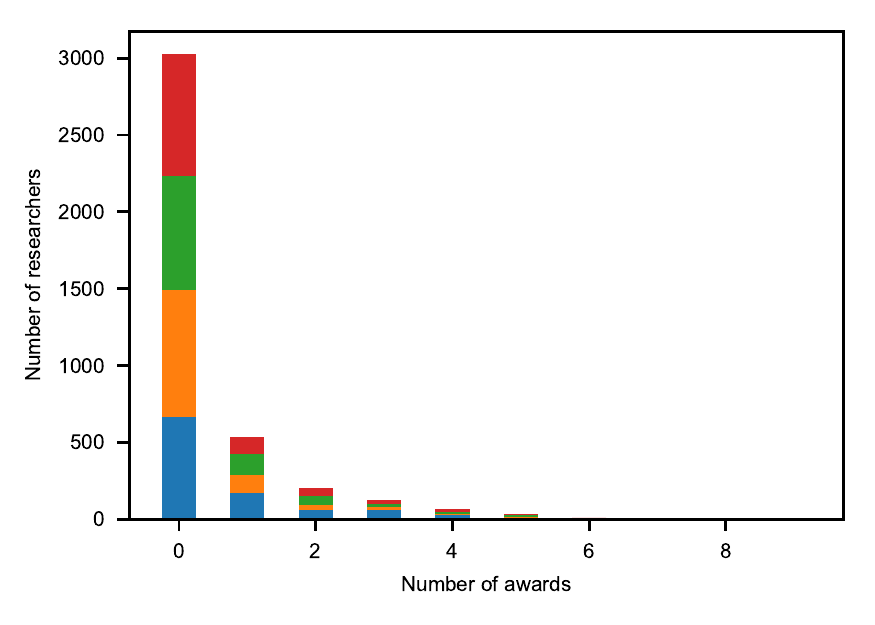}
	\end{subfigure}
	\caption{Award statistics.
		\sflabel{a}~Cumulative number of awards indexed in our data collection.
		\sflabel{b}~Cumulative number of awards to scientists in our datasets.
		\sflabel{c}~Cumulative number of awards to scientists in each research field.
		\sflabel{d}~Distribution of the number of awards garnered by individual scientists.
	}
	\label{fig:awards}
\end{figure}

In total, we trace 1,848 distinct awards to the 4,000 scientists in our dataset. Some scientists have received multiple awards. The number of distinct scientists who have received at least one award in the dataset is 976 (24.4\%). 13.3\% of the researchers received exactly one award, 5.1\% received two, 3.1\% three, and 2.9\% more than three awards (Fig.~\ref{fig:awards_dist}). Of the 1,848 distinct awards, 653 (35.3\%) were granted to researchers in biology, 526 (28.5\%) in economics, 402 (21.8\%) in physics, and 267 (14.4\%) in computer science (Fig.~\ref{fig:awards_yearly}).

\subsection*{Scientometric Measures}
The following paragraphs explain the main scientometric measures that we consider in this work.

\subsubsection*{H-Index}
The h-index, originally proposed by Hirsch in 2005~\cite{Hirsch2005}, is defined as the maximal value of $h$ such that $h$ publications by the author have at least $h$ citations each. Let $N$ be the number of publications and let $\{c_1, \ldots, c_N\}$ the number of citations per paper in decreasing order; i.e.\ $c_i \ge c_j$ for $i < j$. The h-index is given by
\begin{equation}
    \textrm{h} \,=\, \max\left( h \right)
    \quad \textrm{s.t.} \quad
    c_h \ge h
    \,.
\end{equation}

\subsubsection*{C-Index}
We define the c-index as the total number of citations to all publications by the author:
\begin{equation}
	\textrm{c} \,=\, \sum_{i=1}^N c_i \,.
\end{equation}

\subsubsection*{$\mu$-Index}
Lehmann et al.~\cite{Lehmann2006} advocated the use of the mean number of citations per paper:
\begin{equation}
    \mu \,=\, \frac{1}{N} \, \sum_{i=1}^N c_i \,.
\end{equation}

\subsubsection*{G-Index}
Egghe's g-index~\cite{Egghe2006} is a variation on the h-index. It is defined as the maximal value of $g$ such that $g$ publications by the author collectively have at least $g^2$ citations in total:
\begin{equation}
	\textrm{g} \,=\, \max\left( g \right)
	\quad \textrm{s.t.} \quad
	\sum_{i \le g} c_i \,\ge\, g^2
	\,.
\end{equation}

\subsubsection*{O-Index}
The o-index, proposed by Dorogovtsev and Mendes in 2015~\cite{DorogovtsevMendes2015}, is defined as the geometric mean of the h-index (h) and the citation count of the most-cited publication ($c_1$):
\begin{equation}
	\textrm{o} \,=\, \sqrt{\textrm{h} \; c_1} \,.
\end{equation}

\subsubsection*{M-Index}
The m-index, proposed by Bornmann et al.~\cite{Bornmann2008}, is defined as the median number of citations received by publications in the h-core. The h-core comprises the top h publications ranked by citation count. Thus
\begin{equation}
	\textrm{m} \,=\, \textrm{median} \left(  \{ c_1, \ldots, c_\textrm{h} \} \right)  \,.
\end{equation}

\medskip
Based on these traditional scientometric measures, we define their factional counterparts (\emph{-frac}). The fractional measures are based on citation counts $\bar{c}$ that are normalized by the number of authors per publication:
\begin{equation}
	\bar{c} \,=\, \frac{c}{A} \,,
\end{equation}

where $A$ is the number of authors.
The intuition is that this normalization distributes the contribution of a publication equally among the authors. This is clearly a simplification of credit allocation in science~\cite{Waltman2016}, but it is simple and does not introduce new parameters.

\subsubsection*{H-Frac}
The fractional h-index, h-frac, is defined as
\begin{equation}
	\textrm{h-frac} \,=\, \max\left( h \right)
	\quad \textrm{s.t.} \quad
	\bar{c}_h \ge h
	\,.
\end{equation}

Here $\{\bar{c}_1, \ldots, \bar{c}_N\}$ are the normalized citation counts per paper in decreasing order; i.e.\ $\bar{c}_i \ge \bar{c}_j$ for $i < j$.

\subsubsection*{C-Frac}
The fractional measure c-frac is the aggregate of the author's normalized citation counts:
\begin{equation}
	\textrm{c-frac} \,=\, \sum_{i=1}^N \bar{c}_i \,.
\end{equation}

\subsubsection*{$\mu$-Frac}
$\mu$-frac is the mean of the normalized citation counts, averaged over all publications by the author:
\begin{equation}
	\mu\textrm{-frac} \,=\, \frac{1}{N} \, \sum_{i=1}^N \bar{c}_i \,.
\end{equation}

\subsubsection*{G-Frac}
g-frac is likewise defined by analogy with the g-index using normalized citation counts:
\begin{equation}
	\textrm{g-frac} \,=\, \max\left( g \right)
	\quad \textrm{s.t.} \quad
	\sum_{i \le g} \bar{c}_i \,\ge\, g^2
	\,.
\end{equation}

\subsubsection*{O-Frac}
We define o-frac as the geometric mean of the fractional h-index (h-frac) and the largest normalized citation count ($\bar{c}_1$):
\begin{equation}
    \textrm{o-frac} \,=\, \sqrt{\textrm{h-frac} \; \bar{c}_1} \,.
\end{equation}

\subsubsection*{M-Frac}
The fractional counterpart of the m-index, m-frac, is the median of the normalized citation counts among the top h-frac publications ranked by normalized citation counts:
\begin{equation}
    \textrm{m-frac} \,=\, \textrm{median} \left( \{ \bar{c}_1, \ldots, \bar{c}_\textrm{h-frac} \} \right)  \,.
\end{equation}

\subsection*{Effectiveness of Scientometric Measures}

\subsubsection*{ROC Curves}
\begin{figure}
	\centering
	\begin{subfigure}[c]{\textwidth}
		\centering
		\caption{}
		\label{fig:roc}
		\includegraphics[width=\textwidth]{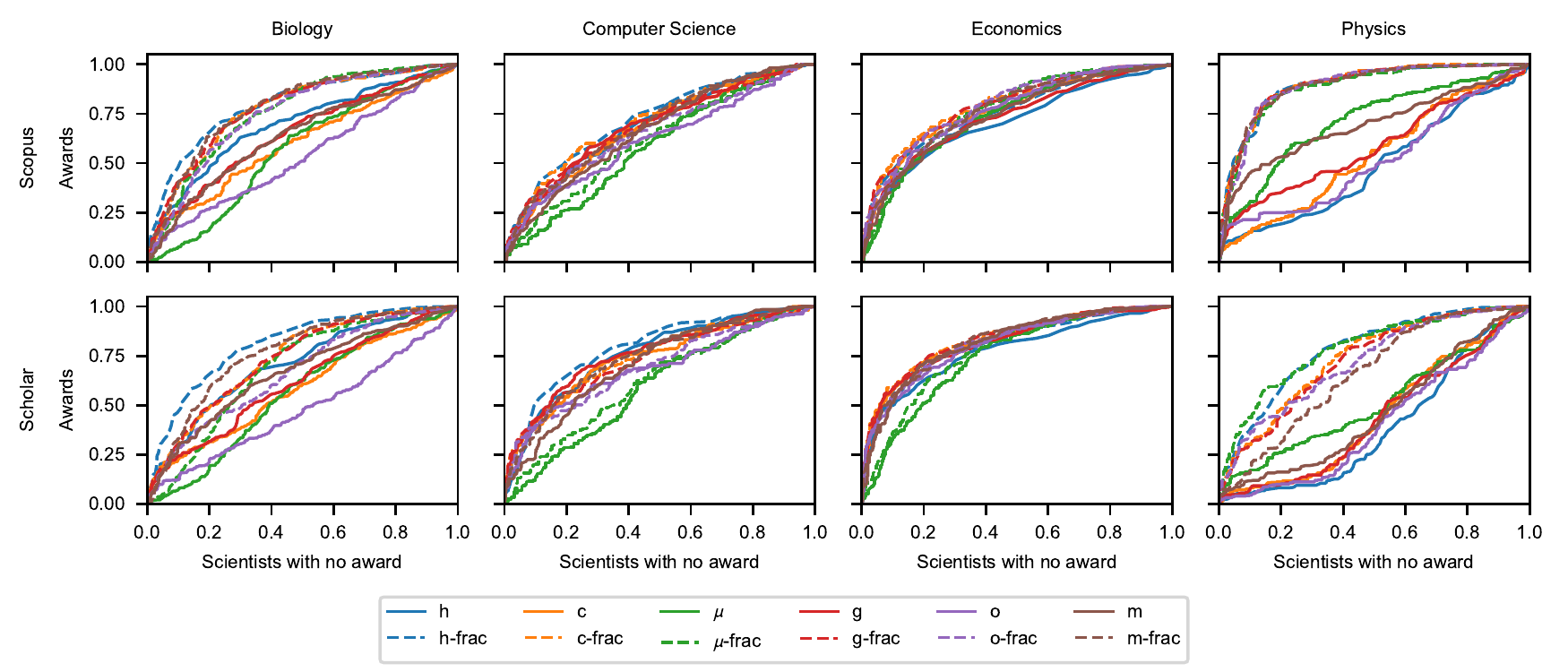}
	\end{subfigure}
	\begin{subfigure}[c]{\textwidth}
		\centering
		\caption{}
		\label{tab:auc}
			\begin{tabu} to 0.9\textwidth{@{\extracolsep{8pt}}X[l]rrrrrrrrX[r]@{}}
	\toprule
	& \multicolumn{4}{c}{Scopus} & \multicolumn{4}{c}{Scholar} & \\
	\cline{2-5} \cline{6-9}
	Measure & Bio & CS & Eco & Phy & Bio & CS & Eco & Phy & Avg.\ \\
	\midrule
c & 0.60 & 0.68 & 0.77 & 0.54 & 0.59 & 0.74 & 0.83 & 0.45 & 0.65 \\
$\mu$ & 0.57 & 0.59 & 0.74 & 0.71 & 0.56 & 0.59 & 0.75 & 0.51 & 0.63 \\
h & 0.69 & 0.69 & 0.71 & 0.49 & 0.70 & 0.76 & 0.78 & 0.39 & 0.65 \\
g & 0.65 & 0.69 & 0.73 & 0.58 & 0.61 & 0.76 & 0.82 & 0.44 & 0.66 \\
m & 0.63 & 0.66 & 0.74 & 0.69 & 0.66 & 0.71 & 0.82 & 0.47 & 0.67 \\
o & 0.53 & 0.62 & 0.77 & 0.52 & 0.48 & 0.68 & 0.81 & 0.41 & 0.60 \\
c-frac & 0.77 & 0.71 & 0.79& \textbf{0.90} & 0.74 & 0.74 & 0.83 & 0.74 & 0.78 \\
$\mu$-frac & 0.76 & 0.62 & 0.77 & 0.88 & 0.68 & 0.62 & 0.77& \textbf{0.79} & 0.74 \\
h-frac& \textbf{0.80}& \textbf{0.72} & 0.76 & 0.89& \textbf{0.81}& \textbf{0.79} & 0.81 & 0.78& \textbf{0.80} \\
g-frac & 0.78 & 0.69 & 0.79 & 0.89 & 0.73 & 0.74& \textbf{0.84} & 0.73 & 0.77 \\
m-frac & 0.78 & 0.69 & 0.77 & 0.89 & 0.76 & 0.74 & 0.82 & 0.66 & 0.77 \\
o-frac & 0.75 & 0.66& \textbf{0.79} & 0.89 & 0.67 & 0.69 & 0.82 & 0.72 & 0.75 \\
\bottomrule
	\end{tabu}
	\end{subfigure}
	\caption{Receiver operating characteristic (ROC) curve and area under the curve (AUC) for each research field and data source.
	\sflabel{a}~The horizontal axis is the accumulated fraction of scientists with no awards (\emph{false positive rate}). The vertical axis is the fraction of awards accumulated by scientists (\emph{true positive rate}). Larger area under the curve (AUC) indicates that a given bibliometric indicator ranks scientists who have received more awards more highly. Details are given in the text.
	\sflabel{b}~Numerical values of AUC for each research field and data source.}
		\label{fig:roc_and_auc}
\end{figure}

We analyze a receiver operating characteristic (ROC) curve for each dataset (Fig.~\ref{fig:roc}).
We rank the scientists by the considered scientometric measure. Lower rank corresponds to higher value of the measure. The scientist with the highest value in the dataset has rank 1. The ROC curve starts at $(0,0)$.
We iterate over the list of scientists, in order of rank $r$ (from 1 onwards), and aggregate the awards. Step $r$ adds the following data point to the ROC curve. The x-coordinate is the fraction of the first $r$ scientists that have not received any award in the dataset (\emph{false positive rate}). The y-coordinate is the fraction of the total number of awards in the dataset received by the first $r$ scientists (\emph{true positive rate}). By construction, the ROC curve ends, for $r=$1,000, at $(1, 1)$. The area under the curve (AUC) is an indicator of the effectiveness of the considered scientometric measure~\cite{Sinatra2016}.
If a measure ranks scientists that have garnered more awards more highly, the ROC curve rises faster and the AUC is higher.

The fractional measures perform much better than their non-fractional counterparts. h-frac performs best across all research areas and datasets (Fig.~\ref{fig:roc_and_auc}).

\medskip
In addition to the AUC, we analyze other criteria that quantify the correlation between a ranking of scientists by a certain scientometric measure and a ranking by the number of awards. If the two rankings are similar (high correlation), the scientometric measure is taken to be a more veridical indicator of scientific reputation.
We evaluate the following correlation measures.

\subsubsection*{Kendall's $\tau$}
We use the $\tau_b$ form of Kendall's $\tau$, which accounts for ties~\cite{Kendall1945}.
It is defined as
\begin{equation}
	\label{eq:kendall}
	\tau
	\;=\; \tau_b \;=\;
	\frac{C - D}{\sqrt{(C + D + T_A) \cdot (C + D + T_B)}} \,,
\end{equation}

where $C$ is the number of concordant and $D$ the number of discordant pairs in two rankings $A$ and $B$. $T_A$ is the number of ties in $A$ only and $T_B$ is the number of ties in $B$ only. If a tie occurs in both $A$ and $B$, it is not added to either $T_A$ or $T_B$.
Equation~\eqref{eq:kendall} reduces to $\tau_a$ when no ties are present~\cite{Kendall1938}:
\begin{equation}
	\tau_a
	\;=\;
	\frac{C - D}{n \, (n-1) / 2} \,,
\end{equation}

where $n$ is the number of elements in $A$ or $B$.

\subsubsection*{Somers' D}
We also measure Somers' D~\cite{Somers1962}. Somers' D of a ranking $A$ w.r.t.\ a ranking $B$ is defined as
\begin{equation}
	\label{eq:somers}
	\textrm{D}_{AB}
	\;=\;
	\frac{\tau_a(A,B)}{\tau_a(B,B)} \,.
\end{equation}

Note that Somers' D is asymmetric. In our evaluation, we set $A$ to the ranking by the considered scientometric measure and $B$ to the ranking based on awards.

\subsubsection*{Goodman and Kruskal's $\gamma$}
Goodman and Kruskal's $\gamma$ is defined as follows~\cite{Goodman1954}:
\begin{equation}
	\label{eq:goodman}
	\gamma
	\;=\;
	\frac{C - D}{C + D} \,.
\end{equation}

\subsubsection*{Spearman's $\rho$}
We also compute Spearman's rank correlation coefficient~\cite{Lovie1995}, which is defined as the Pearson correlation coefficient between the rank variables:
\begin{equation}
	\label{eq:spearman}
	\rho
	\;=\;
	\frac{\textrm{cov}(r_A, r_B)}{\sigma_{r_A} \, \sigma_{r_B}} \,,
\end{equation}

where $r_A$ and $r_B$ are rank variables and $\sigma_{r_A}$ and $\sigma_{r_B}$ the corresponding standard deviations.

\medskip
The results in Table~\ref{tab:effectiveness} support the following observations. First, the fractional measures perform consistently better than their non-fractional counterparts. Furthermore, the relative order of effectiveness of scientometric measures is consistent in the different correlation statistics. This highlights the robustness of our findings. Overall, h-frac is the most effective scientometric measure in terms of correlation with scientific reputation (as indicated by scientific awards).

Of the four research fields we study, economics stands out in terms of the relative effectiveness of different scientometric measures. In economics, g-frac and o-frac appear to be the most effective measures. However, the variation between the scientometric measures in economics is substantially smaller than in the other research fields. For example, the minimal and maximal values of Kendall's $\tau$ in biology in the Scopus dataset are 0.02 and 0.34, while the minimal and maximal values for economics are 0.22 and 0.30 (Table~\ref{tab:effectiveness}(top)). Examination of the data suggests that the field of economics has retained more classical publication patterns, with smaller author sets, fewer publications per author, and minimal hyperauthorship.

\subsection*{Temporal Dynamics}

\subsubsection*{Effectiveness Over Time}
Next, we analyze the effectiveness of scientometric measures in each year from 1990 onwards. To this end, we consider for each year Y the publication and award data up to that year. In particular, we only consider publications up to year Y as well as citations and awards up to the end of year Y. This enables us to investigate the evolution of the effectiveness of scientometric measures over time (Fig.~\ref{fig:effect_and_predict}(top)).

\medskip
We again observe that the fractional measures perform better than their non-fractional counterparts. While most measures tend to decrease in effectiveness over time, the factional measures are more stable. The difference between the fractional and non-fractional measures increases over time. From 2014 onwards, \emph{all} fractional measures are on average more effective than any of the traditional measures (Fig.~\ref{fig:effect_and_predict_summary}(top)). Among all measures, h-frac is the most effective in terms of correlation with scientific reputation (Fig.~\ref{fig:effect_and_predict}(top)).

\subsubsection*{Predictive Power Over Time}
We also investigate the temporal evolution of the predictive power of scientometric measures. Our aim is to quantify how well a scientometric measure predicts \emph{future} scientific reputation.
To this end, we compare a ranking of scientists induced by the considered scientometric measure in year Y to a ranking induced by awards garnered up to year Y + X. A high correlation among these two rankings implies that the scientometric measure is a good predictor of scientific reputation X years into the future.
We compute the same correlation measures defined earlier and take $X=5$ as our default. That is, we measure the ability of scientometric indicators to predict scientific reputation (as evidenced by awards) 5 years in advance (Fig.~\ref{fig:effect_and_predict}(bottom)).

\medskip
Our findings on predictive power are consistent with our earlier findings:
The fractional measures are consistently more predictive than their non-fractional counterparts.
All scientometric measures tend to decline in predictive power over time, but the fractional measures are more stable. The differences between fractional and non-fractional measures increase over time.
From 2014 onwards, all fractional measures are more predictive than the traditional ones (Fig.~\ref{fig:effect_and_predict_summary}(bottom)). h-frac is the most predictive scientometric measure (Fig.~\ref{fig:effect_and_predict}(bottom)).

\clearpage

\begin{figure}
	\centering
	\includegraphics[width=\textwidth]{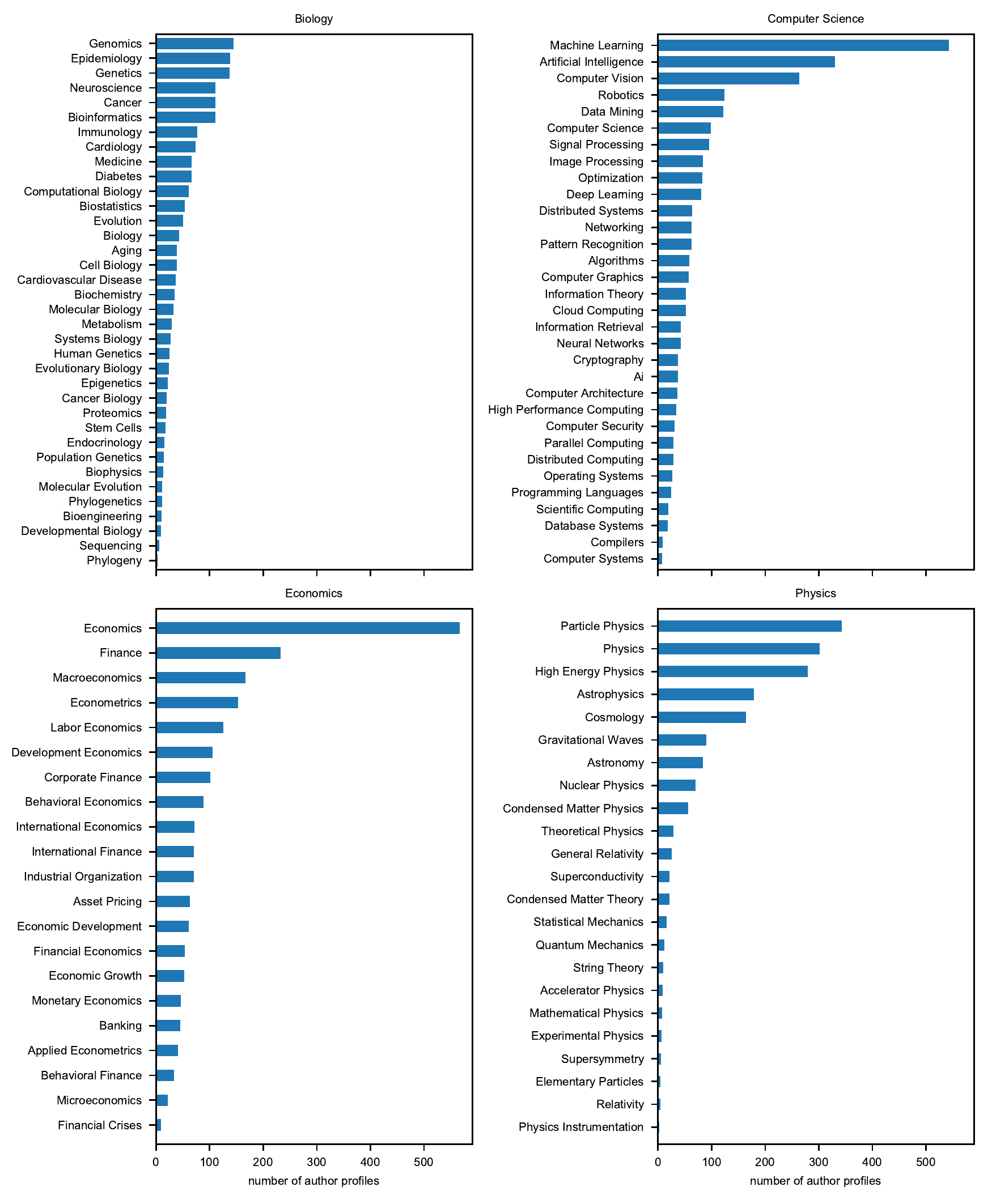}%
	\caption{Google Scholar queries used to initialize the datasets.
		Distribution of search queries in the initial lists of researchers; i.e.\ the number of researchers in the initial lists who feature the respective keyword phrase in their profile.
	}
	\label{fig:scholar_search_labels}
\end{figure}

\clearpage

\begin{figure}
	\centering
	\includegraphics[width=\textwidth]{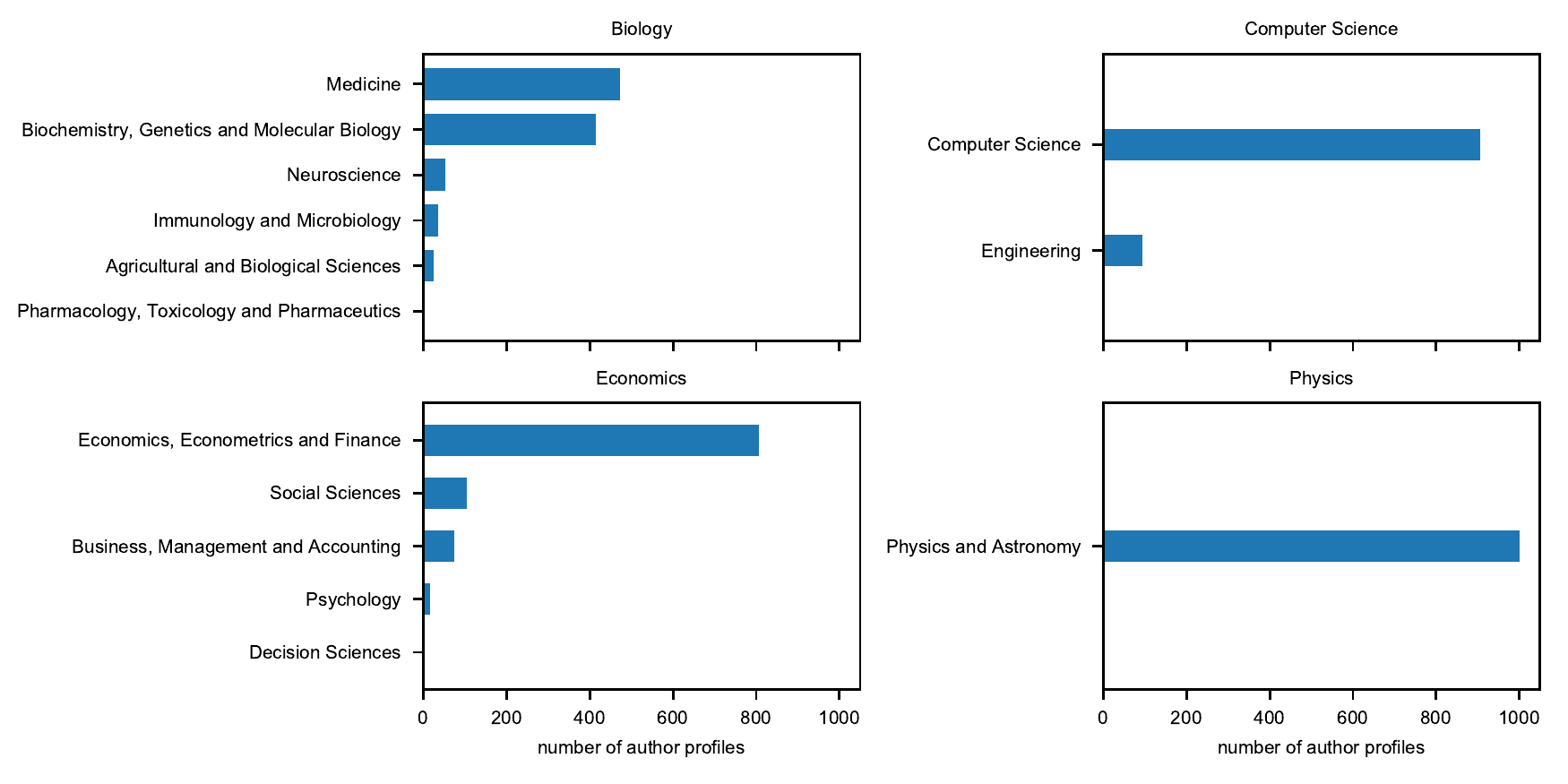}
    \caption{Scopus subject areas used for filtering the initial author list compiled from Google Scholar. The plots show the number of author profiles in the filtered datasets with the respective subject as their primary research area.}
	\label{fig:search_subject_areas}
\end{figure}

\clearpage

\clearpage

\clearpage

\clearpage

\begin{figure}
	\centering
	\includegraphics[width=\textwidth]{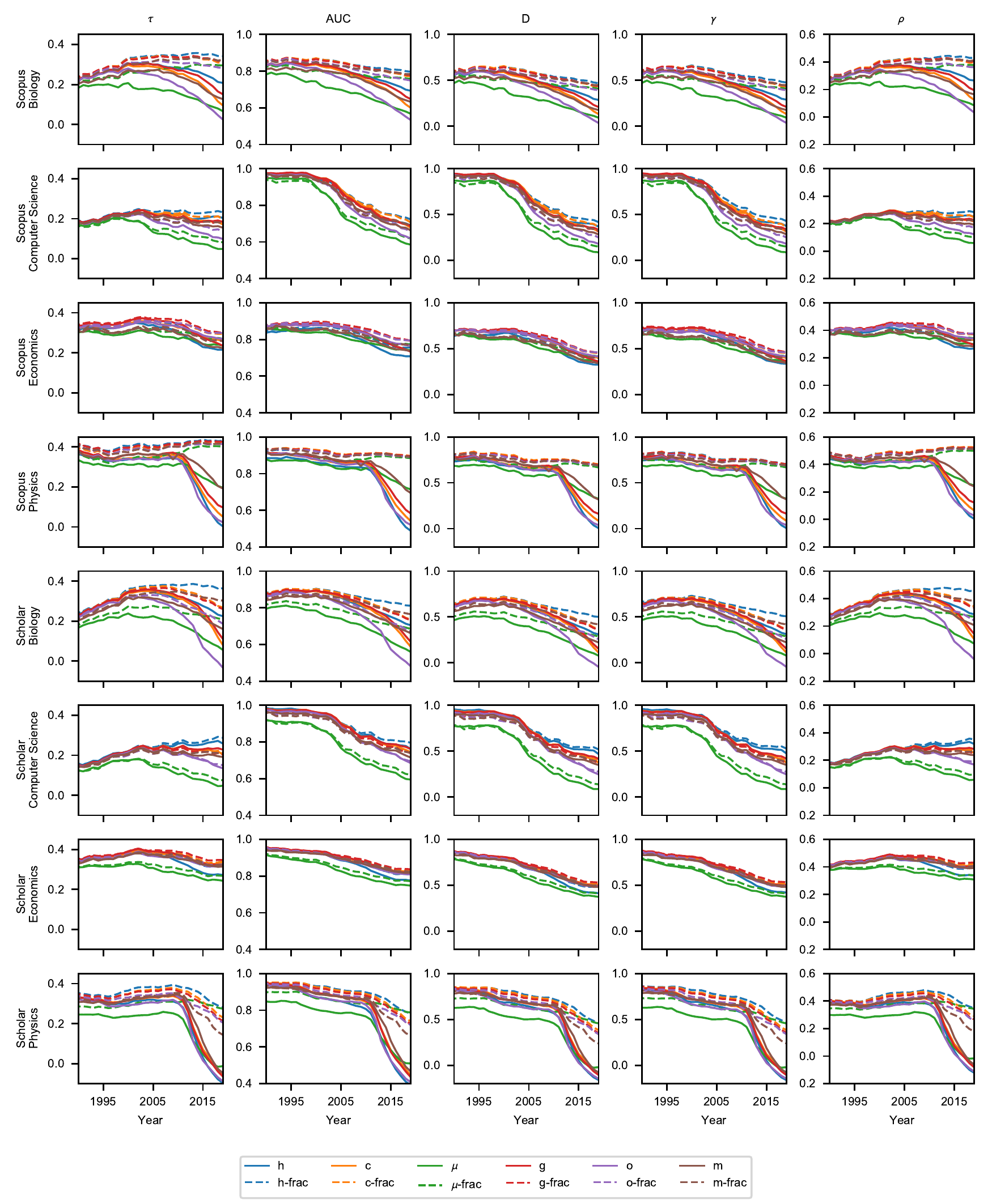}%
	\caption{Effectiveness of scientometric measures over time for different evaluation criteria.
		From left to right:
		Kendall's~$\tau$,
		area under the curve~(AUC),
		Somers'~D,
		Goodman and Kruskal's~$\gamma$,
		Spearman's~$\rho$.
	}
	\label{fig:criteria_all}
\end{figure}

\begin{figure}
	\centering
	\includegraphics[width=\textwidth]{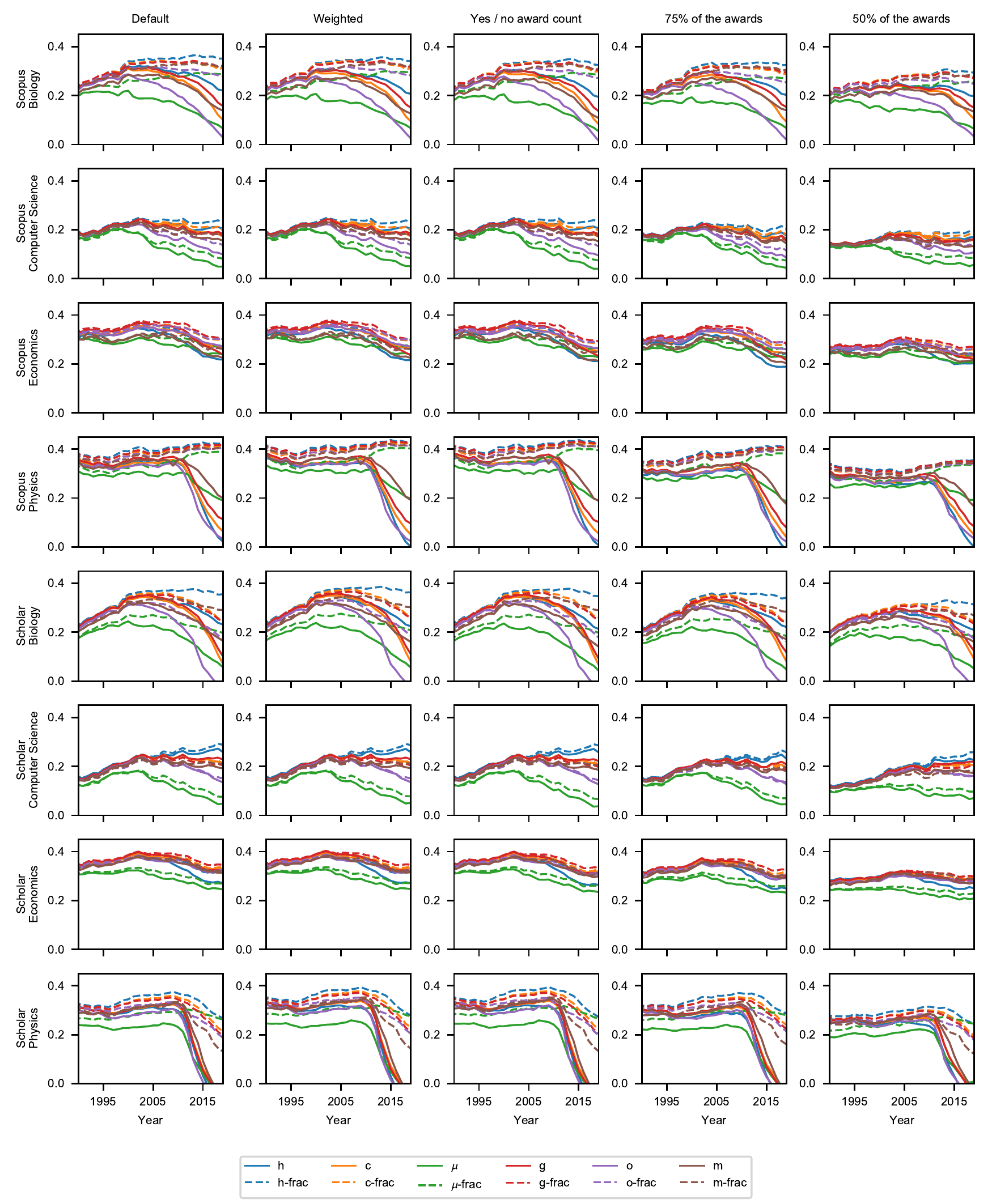}%
	\caption{Effectiveness of scientometric measures over time for different perturbations of rankings induced by awards.
		From left to right:
		equal weight for all awards (default),
		higher weight for awards with $<$ 100 laureates,
		binary (yes / no) award counting,
		random subsets of awards reveiced by researchers in our database (75$\%$ and 50$\%$).
	}
	\label{fig:awards_all}
\end{figure}

\clearpage

\begin{figure}
	\centering
	\includegraphics[width=\textwidth]{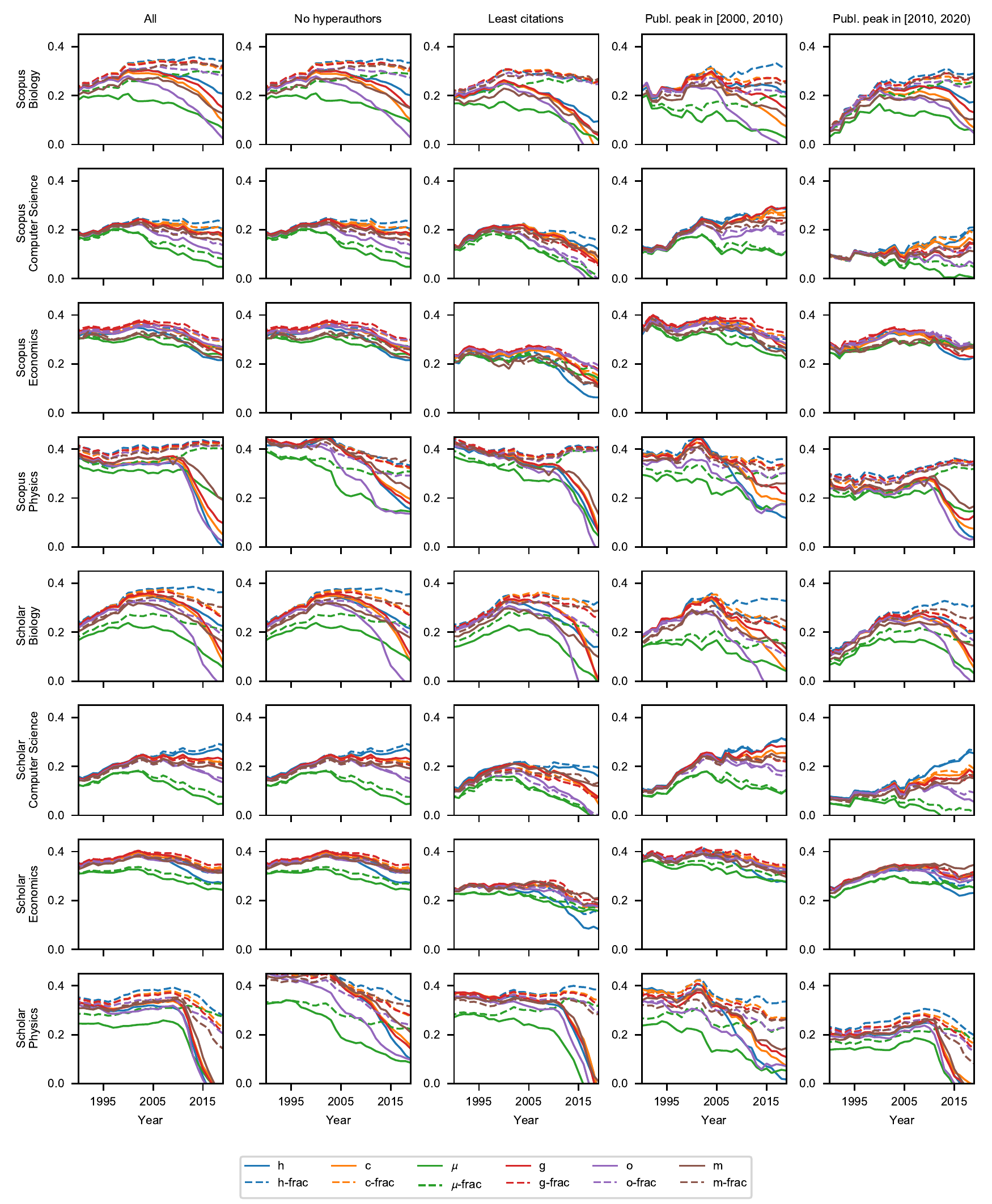}%
	\caption{Effectiveness of scientometric measures over time for different subsets of researchers.
		From left to right:
		all researchers,
		without hyperauthors,
		authors with fewer citations (bottom half),
		authors with publication peak in $\left[2000, 2010\right)$,
		authors with publication peak in $\left[2010, 2020\right)$.
	}
	\label{fig:subsets_all}
\end{figure}

\clearpage

\begin{table}
    \begin{center}
        \caption{Awards used in our study.
			The first five awards apply to all research areas (\emph{cross}-field), while the others are field-specific (\emph{CS} stands for \emph{computer science}). The second-to-last column lists the total number of laureates of each award. The last column shows the number of laureates in our datasets.
			}
		\label{tab:awards}
		\begin{tabularx}{\textwidth}{lXrr}
			\toprule
			& Award & Laureates & Matches \\
			\midrule
			\multirow{5}{*}{\rotatebox{90}{Cross}}
			 & American Academy of Arts \& Sciences & 13,837 & 354 \\
			 & Fellows of the American Association for the Advancement of Science & 65,303 & 390 \\
			 & Fellows of the American Statistical Association & 2,485 & 45 \\
			 & National Academy of Engineering & 4,401 & 106 \\
			 & National Academy of Sciences & 6,085 & 263 \\
			\midrule
			\multirow{4}{*}{\rotatebox{90}{Biology}}
			 & Breakthrough Prize in Life Sciences & 48 & 11 \\
			 & National Academy of Medicine & 2,980 & 121 \\
			 & Nobel Prize in Chemistry & 184 & 5 \\
			 & Nobel Prize in Physiology or Medicine & 219 & 2 \\
			\midrule
			\multirow{2}{*}{\rotatebox{90}{CS}}
			 & ACM Prize in Computing & 13 & 7 \\
			 & Turing Award & 70 & 9 \\
			\midrule
			\multirow{13}{*}{\rotatebox{90}{Economics}}
			 & AEA/AFA Joint Luncheon Speakers & 58 & 10 \\
			 & American Economic Association Distinguished Fellows & 172 & 25 \\
			 & American Economic Association Foreign Honorary Members & 40 & 11 \\
			 & American Economic Association Richard T. Ely Lecturers & 58 & 14 \\
			 & American Finance Association Fischer Black Prize & 8 & 4 \\
			 & Fellows of the American Finance Association & 66 & 22 \\
			 & Fellows of the Econometric Society & 719 & 194 \\
			 & Fisher-Schultz Lecture & 54 & 13 \\
			 & Frisch Memorial Lecture & 9 & 2 \\
			 & John Bates Clark Medal & 42 & 13 \\
			 & Morgan Stanley - AFA Award for Excellence in Finance & 5 & 0 \\
			 & Nobel Prize in Economics & 84 & 17 \\
			 & Walras-Bowley Lecture & 48 & 12 \\
			\midrule
			\multirow{3}{*}{\rotatebox{90}{Physics}}
			 & Breakthrough Prize in Fundamental Physics & 33 & 13 \\
			 & Fellows of the American Physical Society & 10,902 & 178 \\
			 & Nobel Prize in Physics & 213 & 10 \\
			\bottomrule
		\end{tabularx}
	\end{center}
\end{table}

\clearpage

\begin{table}
	\centering
	\caption{Effectiveness of scientometric measures. Higher is better. The most effective measure in each dataset is highlighted in bold.
	}
	\label{tab:effectiveness}
	\begin{tabu} to 0.9\textwidth{@{\extracolsep{8pt}}X[0.2l]X[l]rrrrrrrrX[r]@{}}
	\toprule
	& & \multicolumn{4}{c}{Scopus} & \multicolumn{4}{c}{Scholar} & \\
	\cline{3-6} \cline{7-10}
	& Measure & Bio & CS & Eco & Phy & Bio & CS & Eco & Phy & Avg.\ \\
	\midrule
	\multirow{12}{*}{\rotatebox{90}{Kendall's $\tau$}} &
	c & 0.10 & 0.17 & 0.26 & 0.05 & 0.08 & 0.20 & 0.33 & -0.05 & 0.14 \\
	& $\mu$ & 0.07 & 0.05 & 0.24 & 0.20 & 0.06 & 0.05 & 0.24 & -0.01 & 0.11 \\
	& h & 0.21 & 0.20 & 0.22 & 0.00 & 0.22 & 0.26 & 0.27 & -0.10 & 0.16 \\
	& g & 0.15 & 0.18 & 0.24 & 0.10 & 0.11 & 0.23 & 0.32 & -0.06 & 0.16 \\
	& m & 0.13 & 0.15 & 0.22 & 0.19 & 0.16 & 0.19 & 0.31 & -0.05 & 0.16 \\
	& o & 0.02 & 0.10 & 0.27 & 0.02 & -0.03 & 0.13 & 0.32 & -0.09 & 0.09 \\
	\cline{2-11}
	& c-frac & 0.31 & 0.20 & 0.29 & 0.42 & 0.27 & 0.21 & 0.34 & 0.23 & 0.28 \\
	& $\mu$-frac & 0.30 & 0.08 & 0.27 & 0.40 & 0.21 & 0.08 & 0.27 & 0.27 & 0.23 \\
	& h-frac& \textbf{0.34}& \textbf{0.23} & 0.27& \textbf{0.43}& \textbf{0.36}& \textbf{0.29} & 0.32& \textbf{0.28}& \textbf{0.32} \\
	& g-frac & 0.32 & 0.18 & 0.30 & 0.43 & 0.26 & 0.20& \textbf{0.35} & 0.21 & 0.28 \\
	& m-frac & 0.32 & 0.18 & 0.26 & 0.41 & 0.30 & 0.21 & 0.32 & 0.14 & 0.27 \\
	& o-frac & 0.28 & 0.14& \textbf{0.30} & 0.41 & 0.19 & 0.15 & 0.33 & 0.20 & 0.25 \\
	\midrule
	\multirow{12}{*}{\rotatebox{90}{Somers' D}} &
	c & 0.13 & 0.31 & 0.40 & 0.08 & 0.11 & 0.37 & 0.51 & -0.09 & 0.23 \\
	& $\mu$ & 0.09 & 0.09 & 0.36 & 0.33 & 0.08 & 0.09 & 0.37 & -0.02 & 0.17 \\
	& h & 0.29 & 0.37 & 0.32 & 0.01 & 0.31 & 0.47 & 0.41 & -0.16 & 0.25 \\
	& g & 0.21 & 0.33 & 0.36 & 0.16 & 0.15 & 0.41 & 0.49 & -0.11 & 0.25 \\
	& m & 0.18 & 0.28 & 0.34 & 0.32 & 0.22 & 0.35 & 0.48 & -0.08 & 0.26 \\
	& o & 0.03 & 0.18 & 0.41 & 0.04 & -0.05 & 0.25 & 0.49 & -0.15 & 0.15 \\
	\cline{2-11}
	& c-frac & 0.43 & 0.36 & 0.45& \textbf{0.70} & 0.37 & 0.39 & 0.52 & 0.38 & 0.45 \\
	& $\mu$-frac & 0.41 & 0.15 & 0.41 & 0.67 & 0.29 & 0.14 & 0.41 & 0.46 & 0.37 \\
	& h-frac& \textbf{0.47}& \textbf{0.42} & 0.40 & 0.69& \textbf{0.50}& \textbf{0.52} & 0.49& \textbf{0.46}& \textbf{0.49} \\
	& g-frac & 0.44 & 0.32 & 0.45 & 0.69 & 0.36 & 0.37& \textbf{0.53} & 0.35 & 0.44 \\
	& m-frac & 0.45 & 0.33 & 0.40 & 0.69 & 0.42 & 0.39 & 0.50 & 0.24 & 0.42 \\
	& o-frac & 0.39 & 0.25& \textbf{0.46} & 0.69 & 0.26 & 0.27 & 0.50 & 0.34 & 0.39 \\
	\midrule
	\multirow{12}{*}{\rotatebox{90}{Goodman and Kruskal's $\gamma$}} &
	c & 0.13 & 0.31 & 0.40 & 0.08 & 0.11 & 0.37 & 0.51 & -0.09 & 0.23 \\
	& $\mu$ & 0.09 & 0.09 & 0.36 & 0.33 & 0.08 & 0.09 & 0.37 & -0.02 & 0.17 \\
	& h & 0.29 & 0.37 & 0.34 & 0.01 & 0.31 & 0.47 & 0.42 & -0.16 & 0.26 \\
	& g & 0.21 & 0.34 & 0.36 & 0.16 & 0.15 & 0.42 & 0.50 & -0.11 & 0.25 \\
	& m & 0.18 & 0.28 & 0.34 & 0.32 & 0.22 & 0.35 & 0.48 & -0.08 & 0.26 \\
	& o & 0.03 & 0.18 & 0.41 & 0.04 & -0.05 & 0.25 & 0.49 & -0.15 & 0.15 \\
	\cline{2-11}
	& c-frac & 0.43 & 0.36 & 0.45 & 0.70 & 0.37 & 0.39 & 0.52 & 0.38 & 0.45 \\
	& $\mu$-frac & 0.41 & 0.15 & 0.41 & 0.67 & 0.29 & 0.14 & 0.41 & 0.46 & 0.37 \\
	& h-frac& \textbf{0.48}& \textbf{0.43} & 0.42& \textbf{0.70}& \textbf{0.51}& \textbf{0.53} & 0.50& \textbf{0.47}& \textbf{0.50} \\
	& g-frac & 0.44 & 0.32& \textbf{0.46} & 0.70 & 0.36 & 0.37& \textbf{0.53} & 0.35 & 0.44 \\
	& m-frac & 0.45 & 0.33 & 0.40 & 0.69 & 0.42 & 0.39 & 0.50 & 0.24 & 0.42 \\
	& o-frac & 0.39 & 0.25 & 0.46 & 0.69 & 0.26 & 0.27 & 0.50 & 0.34 & 0.39 \\
	\midrule
	\multirow{12}{*}{\rotatebox{90}{Spearman's $\rho$}} &
	c & 0.12 & 0.21 & 0.33 & 0.06 & 0.10 & 0.25 & 0.41 & -0.07 & 0.18 \\
	& $\mu$ & 0.08 & 0.06 & 0.30 & 0.25 & 0.07 & 0.06 & 0.31 & -0.02 & 0.14 \\
	& h & 0.26 & 0.25 & 0.27 & 0.00 & 0.29 & 0.32 & 0.34 & -0.12 & 0.20 \\
	& g & 0.19 & 0.23 & 0.29 & 0.12 & 0.14 & 0.28 & 0.40 & -0.08 & 0.20 \\
	& m & 0.16 & 0.19 & 0.28 & 0.24 & 0.21 & 0.24 & 0.39 & -0.06 & 0.21 \\
	& o & 0.03 & 0.12 & 0.34 & 0.03 & -0.04 & 0.17 & 0.40 & -0.11 & 0.12 \\
	\cline{2-11}
	& c-frac & 0.40 & 0.24 & 0.37& \textbf{0.53} & 0.34 & 0.26 & 0.42 & 0.29 & 0.36 \\
	& $\mu$-frac & 0.38 & 0.10 & 0.34 & 0.50 & 0.27 & 0.09 & 0.34 & 0.34 & 0.30 \\
	& h-frac& \textbf{0.43}& \textbf{0.28} & 0.33 & 0.52& \textbf{0.46}& \textbf{0.35} & 0.40& \textbf{0.35}& \textbf{0.39} \\
	& g-frac & 0.40 & 0.22 & 0.37 & 0.52 & 0.33 & 0.25& \textbf{0.43} & 0.27 & 0.35 \\
	& m-frac & 0.41 & 0.22 & 0.33 & 0.51 & 0.38 & 0.26 & 0.41 & 0.18 & 0.34 \\
	& o-frac & 0.36 & 0.17& \textbf{0.38} & 0.51 & 0.24 & 0.19 & 0.41 & 0.26 & 0.31 \\
	\bottomrule
	\end{tabu}
\end{table}

\clearpage

    \printbibliography[title=Additional References,segment=\therefsegment,filter=notother]

\end{refsegment}

\end{document}